\PassOptionsToPackage{table,xcdraw,dvipsnames}{xcolor}
 \documentclass[sigconf, table, natbib=true]{acmart}

\usepackage{embedfile}

\usepackage[utf8]{inputenc}
\makeatletter%
\@ifclassloaded{acmart}{
  \makeatletter
  \def\mdseries@tt{m}
  \makeatother
}{}
\makeatother

\usepackage{enumitem}

\usepackage{epsfig}
\usepackage{wrapfig}
\usepackage{pgfplotstable}
\usepackage{pgfplots}
\usepgfplotslibrary{groupplots}
\pgfplotsset{compat=newest} %
\usepgfplotslibrary{dateplot}
\usepackage{amsmath,mathtools}
\usepackage{amssymb}
\usepackage{amsfonts}
\usepackage{mathrsfs}
\usepackage{relsize}
\usepackage{multicol}
\usepackage{bbm}
\usepackage{listings}
\usepackage{float}
\usepackage{microtype}
\usepackage{tikz}
\usetikzlibrary{decorations.text,calc,shapes,arrows,arrows.meta, positioning,shapes.misc,decorations.markings,decorations.markings,decorations.pathreplacing,matrix,spy}
\usetikzlibrary{fadings}
\usetikzlibrary{patterns}
\usetikzlibrary{shadows.blur}

\usepackage{svg}

\usepackage[]{appendix}

\usepackage{booktabs}
\usepackage{multirow}
\usepackage{makecell}

\usepackage{adjustbox}
\usepackage{verbatim}
\usepackage[T1]{fontenc}
\usepackage{adjustbox} %
\usepackage{graphicx}
\usepackage{caption}
\usepackage[noend]{algpseudocode}
\usepackage{algorithm}

\usepackage{siunitx}
\usepackage{nicefrac}
\usepackage[colorinlistoftodos]{todonotes}
\usepackage{spverbatim}
\usepackage{seqsplit}

\newcommand{\mcrot}[4]{\multicolumn{#1}{#2}{\rlap{\rotatebox{#3}{#4}~}}} 

\usepackage{array}
\newcolumntype{R}[2]{%
    >{\adjustbox{angle=#1,lap=\width-(#2)}\bgroup}%
    l%
    <{\egroup}%
}

\newcommand*\textcite[1]{\citet{#1}}

\usepackage[autostyle, english=american]{csquotes}
\MakeOuterQuote{"}

\makeatletter
\pgfplotsset{
    every axis x label/.append style={
        alias=current axis xlabel
    },
    legend pos/outer south/.style={
        /pgfplots/legend style={
            at={%
                (%
                \@ifundefined{pgf@sh@ns@current axis xlabel}%
                {xticklabel cs:0.5}%
                {current axis xlabel.south}%
                )%
            },
            anchor=north
        }
    }
}
\makeatother

\newcolumntype{t}{>{\ttfamily}l}
\newcolumntype{T}{>{\ttfamily}c}

\newcolumntype{$}{>{\global\let\currentrowstyle\relax}}
\newcolumntype{^}{>{\currentrowstyle}}

\DeclareMathOperator*{\argmax}{arg\,max}

\everypar{\looseness=-1} %
\begin{document}

\copyrightyear{2020} 
\acmYear{2020} 
\setcopyright{licensedusgovmixed}\acmConference[AISec'20]{13th ACM Workshop on Artificial Intelligence and Security}{November 13, 2020}{Virtual Event, USA}
\acmBooktitle{13th ACM Workshop on Artificial Intelligence and Security (AISec'20), November 13, 2020, Virtual Event, USA}
\acmPrice{15.00}
\acmDOI{10.1145/3411508.3421372}
\acmISBN{978-1-4503-8094-2/20/11}

\title{Automatic Yara Rule Generation Using Biclustering}

  \author{Edward Raff}
  \orcid{0000-0002-9900-1972}
    \affiliation{%
    \institution{Laboratory for Physical Sciences}
  }
  \affiliation{%
    \institution{Booz Allen Hamilton}
  }
  \email{ edraff@lps.umd.edu}
  
  \author{Richard Zak}
  \orcid{0000-0003-4272-2565}
    \affiliation{%
    \institution{Laboratory for Physical Sciences}
  }
  \affiliation{%
    \institution{Booz Allen Hamilton}
  }
\email{ rzak@lps.umd.edu}
  
  \author{Gary Lopez Munoz}
  \affiliation{%
    \institution{Laboratory for Physical Sciences}
  }
  \affiliation{%
    \institution{Booz Allen Hamilton}
  }
  \email{dlmgary@lps.umd.edu}

    \author{William Fleming}
    \affiliation{%
    \institution{U.S. Navy}
  }
  \email{ william.r.fleming1@navy.mil}
  
  \author{Hyrum S. Anderson}
  \authornote{Work done while at Elastic.}
  \affiliation{%
    \institution{Microsoft}
  }
  \email{hyruma@microsoft.com}
  
  \author{Bobby Filar}
  \affiliation{%
    \institution{Elastic NV}
  }
  \email{robert.filar@elastic.co}
  
    \author{Charles Nicholas}
  \affiliation{%
    \institution{Univ. of Maryland, Baltimore County}
  }
  \email{nicholas@umbc.edu}
  
  \author{James Holt}
  \affiliation{%
    \institution{Laboratory for Physical Sciences}
  }
  \email{ holt@lps.umd.edu}

\begin{abstract}
Yara rules are a ubiquitous tool among cybersecurity 
practitioners and analysts. Developing high-quality Yara rules to detect a malware family of interest can be labor- and time-intensive, even for expert users. Few tools exist and relatively little work has been done on how to automate the generation of Yara rules for specific families. In this paper, we leverage large n-grams ($n \geq 8$) combined with a new biclustering algorithm to construct simple Yara rules more effectively than currently available software. Our method, AutoYara, is fast, allowing for deployment on low-resource equipment for teams that deploy to remote networks. 
Our results demonstrate that AutoYara can help reduce analyst workload by producing rules with useful true-positive rates while maintaining low false-positive rates, sometimes matching or even outperforming human analysts.
In addition, real-world testing by malware analysts indicates AutoYara could reduce analyst time spent constructing Yara rules by 44-86\%, allowing them to spend their time on the more advanced malware that current tools can't handle. Code will be made available at
\url{https://github.com/NeuromorphicComputationResearchProgram}.
\end{abstract}

\settopmatter{printfolios=true} %
\maketitle
\renewcommand{\shortauthors}{Raff et al.}

\section{Introduction}

Machine Learning has become more involved in malware detection systems and cybersecurity, but older signature-based approaches are still an important tool. In particular, Yara \cite{Alvarez2013} is widely used to specify signatures and perform searches. Yara is a tool to combine content matching against simple regular expressions with logic rules, and these rules `fire' if the predicates are satisfied. These predicates combined are often called `Yara Rules', and may be used to identify specific malware families, the presence of CVEs, specific signatures of functionality, or generic indicators of maliciousness. 

Developing effective Yara rules can be very time intensive, especially for junior analysts who lack deeper expertise and intuition on what should be included in a Yara rule to achieve a goal. For example, a related task in performing reverse engineering (a task that may be necessary to build good signatures for difficult malware samples) can take several hours to weeks for a single file, even for expert users with over a decade of experience \cite{Votipka2019}. In our experience, analysts rarely get through all of their "necessary" tasks and work under a continually growing backlog of samples that need to be analyzed or have rules created. 
Despite Yara's widespread use, only a few works have attempted to automate the development of Yara rules. 

We consider the problem of trying to develop a Yara rule to identify a specific malware family given only a limited number of example files from that family. A common workflow in developing Yara rules is to manually inspect multiple files to determine common contents, wrapped by trial-and-error refinement of the developed rules, where success is measured against coverage and false positive rates on a collection that includes benign or out-of-family files. 

In this paper, we are concerned with two practical use-cases. First, a "hunt" team is deployed to an unfamiliar network after the discovery of malicious files. To determine the extent of the attack they craft Yara rules to perform a broader search across the network. In this scenario, there are two primary concerns: 1) Yara rules that generate a lot of false positives (e.g. returning a significant amount of benign files) could slow the investigation and 2) security workers often have fewer ($\leq 10$) samples when creating a Yara rule. 
The second scenario is based on scaling Yara rule construction to track specific malware families that are known to be difficult to correctly classify, due to their structural resemblance to benign software, or because they are polymorphic in nature.
We test this scenario on live production data to demonstrate that our approach could save analysts %
significant time spent constructing Yara rules. 

We stress that our objective is not to entirely replace a human analyst in producing Yara rules. A skilled analyst will likely be able to produce better rules than our tool given the time, and techniques such as packing will successfully thwart our tools. The goal is to provide a tool that can produce rules that are good enough that they can often be used without alteration, or quickly improved so that analysts can get through their workload faster. Given a satisfying AutoYara result, analysts can spend their limited and valuable time working on more challenging samples and tasks that do not yet yield to automation.

\begin{figure}[!htb]

\begin{lstlisting}[breaklines=true,basicstyle=\footnotesize]
rule Analyst {
    strings:
     $a = "191231235959Z0U1", $b = "downloader" wide
     $c = "1.0.2.417" wide
    condition:
     $a and $b and $c
}
rule YarGen {
 strings:
  $s1 = "5054585<5@5D5H5T5X5\\5`5d5h5" fullword ascii
  $s2 = "0 0$0(0,0004080D0T0X0d0h0l0(6" fullword ascii
  ...
  $s20 = "</<B<P<T<X<\\<`<d<h<" fullword ascii
 condition:
  ( uint16(0) == 0x5a4d and
     filesize < 1000KB and
     pe.imphash() == "50542991fab95f9ee910f48e0fd7f114"
     and ( 8 of ($s*) )
  ) or ( all of them )
}
rule AutoYara{
  strings:
    $x1 =  { 07 A2 4F E5 87 ... DC EC }
    ...
    $x31 = { CF E8 A5 30 57 ... B0 8E }
  condition:
    (20 of ($x0,$x1,...,$x26)) or (7 of ($x27,...,$x35))
}
\end{lstlisting}

\caption{Examples of Yara rules generated by 1) an analyst, 2) YarGen and 3) our AutoYara.   } \label{fig:yara_examples}
\end{figure}

Examples of Yara rules generated by an analyst, a prior method called YarGen, and our new AutoYara are shown in \autoref{fig:yara_examples}. Analysts can use their expertise to select the salient factors to make compact and effective rules. Prior tools like YarGen rely on a number of heuristics and string features and have varying levels of effectiveness. Our new approach makes larger rules, but uses the redundancy and conjunction of components to achieve the extremely low false-positive rates that analysts desire. 

We perform this task in two steps. First, we leverage recent work in  finding frequent larger n-grams, for $n \leq 1024$, to find several candidate byte strings that could become features. Second, we extend the SpectralCoClustering algorithm to work when the number of biclusters is not known \textit{a priori}. We will show that biclustering allows us to easily produce complex logic rules that allow us to build signatures with low false positive rates, e.g. $\leq 0.1\%$. Third, we perform the first production comparison of signature generation against professional analysts, giving us a benchmark for human ability and the relative effectiveness of the AutoYara tool. 

We review the related work on automated Yara rule construction for Windows PE executables in \autoref{sec:related_work}. The intuition on how we will use biclustering, and our improved method for it, is given in \autoref{sec:bicluster}.  Next, we discuss the design of our AutoYara system in \autoref{sec:autoyara}, which uses our bicluster algorithm with only a few simple pruning steps to build a Yara rule. We review the datasets used for building AutoYara in \autoref{sec:datasets}. We perform extensive and real-world evaluations in \autoref{sec:results}, where we see that AutoYara can be effective for the scenarios in which we are interested. 
\autoref{sec:autoyara_depth} presents an initial investigation into AutoYara's results compared to a human analyst. 
Finally, we summarize the approach and results in \autoref{sec:conclusion}. 

\section{Related Work}\label{sec:related_work}

Yara \cite{Alvarez2013} is an industry standard regular expression tool designed for malware analysis. Many malware analysis tools support Yara directly, so we want to look at methods which also create compatible signatures. 
The only other tools that are readily available and actively maintained for generating Yara rules from example files are YarGen \cite{Roth2013} and VxSig \cite{Blichmann2008}, which is considered state-of-the-art. Other works have developed malware family classification models and called these approaches "signatures" as well (e.g., \cite{David2015}). These systems do not produce Yara rules, however, so we do not consider them viable alternatives.

YarGen uses a Naive Bayes model to score the potential utility of features that can be extracted from a binary, predominately strings. Then YarGen uses a number of heuristics to select the features to use, and combines them in a heuristic fashion. The authors of YarGen encourage its use as a starting point for rule construction, and to build rules by manually adjusting and refining YarGen's output. This recommendation aligns with our experience, where YarGen often produces rules with 0 hits, and requires alteration to be useful. 

The other existing and available approach was proposed by \textcite{Blichmann2008}, which uses a least-common-subsequence (LCS) algorithm to find byte sequences, extracted from functions, that appear to be common to all files in the given sample. This constrains their rules to only code, where our approach can leverage information found from all regions of a binary. We compare against the VxSig\footnote{\url{https://github.com/google/vxsig}} implementation of this approach. While a powerful tool, we found in practice that 
VxSig required human intervention to make it work, and could take hours per malware sample to run. 
Details about the approach we used with VxSig to increase the rate of successful rule production can be found in \autoref{sec:vxsig}.

There are a number of prior works that could be described as "Greedy" rule construction, which also leveraged large $n \in [32, 256]$ byte grams to create rules from an input. These works have generally pursued a strategy of collecting a large set of malware ($\geq 10,000$ samples ) that corresponds to multiple families, and sought to create a set of rules to cover the maximum set of input files with few false positives \cite{Kim2004,Newsome:2005:PAG:1058433.1059393}.  These greedy construction approaches have often used discriminative models like Naive Bayes --- similar to YarGen \cite{Yegneswaran:2005:AGS:1251398.1251405}. Signatures were constructed by greedily adding byte-grams based on a number of different strategies.  The most recent work by \textcite{Griffin2009a} required 33 GB of benign data available to cover 48k malicious files, with 48-gram features needing a 17.5GB Markov language model and another 52 GB for a search index. We have implemented a representative greedy strategy in this work, which we will refer to as the "Greedy" method. To make this Greedy method run in a reasonable amount of time, we have replaced the larger multi-GB indexes with the KiloGram approach. The Greedy approaches do not work well in our situation because we are trying to build rules for a specific family with as few as two examples. Our results also differ in that we develop a biclustering based strategy to build complex logic rules, that we found more effective than greedy search. 

For computational efficiency, our AutoYara system leverages the recently proposed work in "KiloGrams" that demonstrates how to find the top-$k$ most frequent $n$-grams in a computationally efficient manner with fixed memory cost~\cite{Kilograms_2019,hashgram_2018,raff_hash_gram_parallel}. Their work argued that these large byte grams can be used as Yara rules. However, their approach could not be deployed operationally, as it requires having all training data available and building a new model from scratch for every Yara rule, and an even distribution of malware samples between all families. This is impractical. 

We note that others have elucidated a number of fundamental limits on the potential effectiveness of classical signatures and attempts to learn signatures \cite{Song:2007:IMP:1315245.1315312,Newsome:2005:PAG:1058433.1059393}, and that learning signatures can be subject to adversarial attacks that hamper signature creation \cite{Perdisci2006,Newsome:2006:PTS:2166373.2166380}. Indeed, approaches like packing can easily thwart the Yara signatures we wish to generate. We are not challenging or circumventing any of these fundamental limits. They are a weakness of \textit{any} signature based approach. But since signatures are still a widely used tool, we believe there is value in improving these processes with the understanding that such tools are useful, if not perfect.

\section{Improved Spectral CoClustering } \label{sec:bicluster}

\begin{figure}[!ht]
    \centering
    \adjustbox{width=1.0\columnwidth}{%
    \includegraphics[width=\columnwidth]{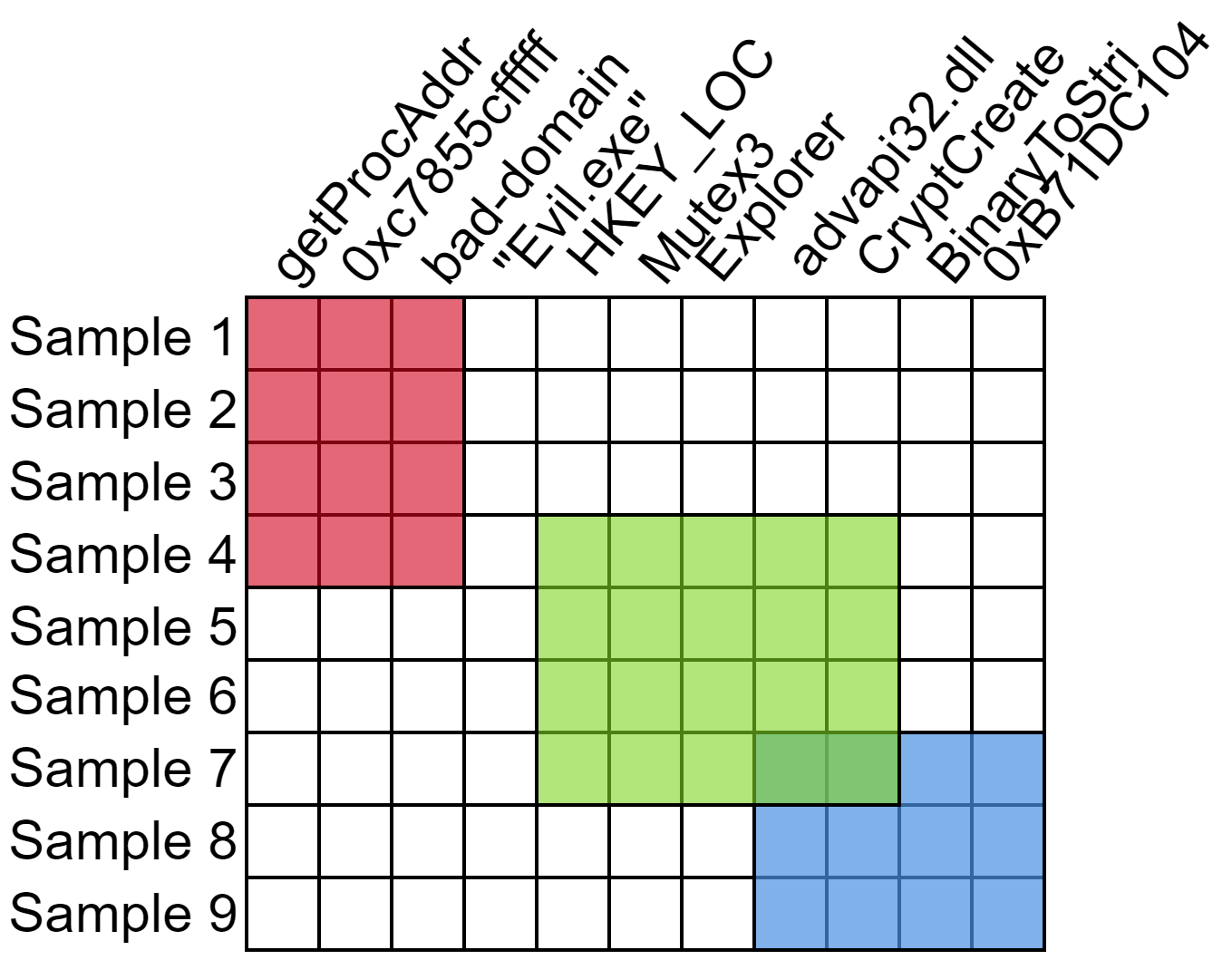}
    }
    \caption{Illustration of the type of biclustering we wish to find. It tells us that three groups of features (red, green, \& blue) could be useful to create a signature that covers all 9 samples. It also identifies that the "Evil.exe" feature would not be useful in this case. We do not care if biclusters overlap, as multiple clauses can use the same features.  }
    \label{fig:bicluster}
\end{figure}

To create our Yara rule signatures, we will make use of biclustering algorithms as the fundamental mechanism of rule construction. The goal of bi-cluster is to take an input matrix $X$, and to simultaneously cluster the rows and columns of the matrix to reveal an underlying structure between columns and rows \cite{Hartigan1972}. 
An example of a biclustering is given in \autoref{fig:bicluster}. 

In Figure~\ref{algo:bicluster} we see that three bi-clusters were found. In general, biclusters may overlap in rows or columns, and several rows and/or columns may not belong to any bicluster. We are interested in biclustering because it gives us a natural means to extract Yara rules. To build a good Yara rule, we want to know 1) which features should be used at all, and 2) which features should be combined into an "and" statement (would reduce FPs), and which should be placed into an "or" statement (would increase TPs). 

Our first contribution is that biclustering gives us a simple approach to do this jointly over the features, rather than considering greedy approaches that select features one at a time. For every feature within a bicluster, we \texttt{<and>} them together since they co-occur, and we \texttt{<or>} the predicates built from biclusters.  This results in a ``disjunction of conjunctions''
rule formulation, where, referring back to \autoref{fig:bicluster}, we build the rule over the features $F_i$ as $(F_1 \land  F_2 \land  F_3) \lor (F_5 \land F_6 \land F_7 \land F_8) \lor (F_7 \land F_8 \land F_9 \land F_10)$. In this way can build complex rules with multiple terms, with sharing of terms between clauses.

The difficulty is in performing the biclustering process itself. In particular, the majority of biclustering algorithms  require specifying the number of biclusters in advance \cite{Role2019}, which is unknown in our situation, and enforce no overlap between biclusters. We desire a biclustering method that can determine the number of biclusters automatically, even if it is only one bicluster, allows overlapping biclusters, and will discard rows and columns that do not fit in any bicluster. 

We develop this by extending the seminal SpectralCoClustering approach of \cite{Dhillon:2001:CDW:502512.502550}. Their approach is widely used for both its simplicity and effectiveness. By computing the Singular Value Decomposition (SVD) of a normalized input matrix $A \in \mathbb{R}^{r, c}$, they create a new matrix $Z \in \mathbb{R}^{r+c,\log_2 k}$, where the first $r$ rows of $Z$ correspond to the original $r$ rows of $A$, and the remaining rows of $Z$ correspond to the columns of $A$. The number of features in the transformed matrix is $\log_2 k$, where $k$ is the desired number of biclusters. The biclusters are then found by running the $k$-means algorithm on $Z$, and the rows of $Z$ found in the cluster tell us which rows/columns of $A$ are in the final biclustering.  \cite{Dhillon:2001:CDW:502512.502550} proved that this corresponds to a weighted cut of the bipartite graph to perform biclustering.

\begin{algorithm}[!ht]
\caption{Adaptive SpectralCoClustering} \label{algo:bicluster}
\begin{algorithmic}[1]
\Require Set of files/data points $\mathcal{S}$, and set of n-gram features $\mathcal{G}$
\State construct matrix $\boldsymbol{A} \in \mathbb{R}^{r,c}$, where $r = |\mathcal{S}|$ is the number of data points and $c=|\mathcal{G}|$ is the number of columns / n-gram features 
\State $R_{i,i} = \sum_{j=1}^c \tilde{A}_{i,j}$
\State $C_{j,j} = \sum_{i=1}^r \tilde{A}_{i,j}$
\State Compute normalized matrix $A_{n}=R^{-1 / 2} A C^{-1 / 2}$
\State Set max SVD dimensions $\ell \gets  \log_2 \left( \min(r, c)/2 \right)$
\Comment{The following two lines show 'Scale' based normalization. One could also use bistochastic or log-based normalization proposed in \cite{Kluger2003}}
\State $U, S, V \gets $ ThinSVD($A_n$, $\ell$+1)
\State $Z \gets \left[\begin{array}{l}
{R^{-1 / 2} U} \\
{C^{-1 / 2} V}
\end{array}\right]$, a new dataset in $\mathbb{R}^{r+c,\ell}$
\State $\mu_1, \ldots, \mu_k$, $\Sigma_1, \ldots, \Sigma_k$ $\gets $  Variational GMM \cite{Rasmussen2000,export:67239} clustering results on the $r+c$ rows of $Z$ 
\State $\mathcal{B} \gets $ empty set of bi-clusters
\ForAll{GMM clusters $\mu_i, \Sigma_i$}
    \State $\alpha_i \gets $ all rows $j$ of $Z$ s.t. $P\left(z_k \mid	 \mathcal{N}(\mu_i, \Sigma_i)\right) > 1/(k+1)$
\EndFor
\ForAll{$\alpha_i$} \Comment{Filter out biclusters that contain essentially only rows from $\mathcal{S}$ or only columns from $\mathcal{G}$  }
    \If{$\sum_{\forall j \in \alpha_i} \mathbbm{1}[j \leq r] \leq 1$ or $\sum_{\forall j \in \alpha_i} \mathbbm{1}[j > r] \leq 1$}
        \State discard/remove cluster $\alpha_i$
    \EndIf
\EndFor
\State $c_{min} \gets \min\left(5, \argmax_i \sum_{\forall j \in \alpha_i} \mathbbm{1}[j > r] \right)$
\State $r_{min} \gets \min\left(5, \argmax_i \sum_{\forall j \in \alpha_i} \mathbbm{1}[j \leq r] \right)$
\ForAll{$\alpha_i$}
    \If{$\sum_{\forall j \in \alpha_i} \mathbbm{1}[j \leq r] < r_{min}$ or $\sum_{\forall j \in \alpha_i} \mathbbm{1}[j > r] < c_{min}$}
        \State discard $\alpha_i$
    \EndIf
\EndFor
\State \textbf{return} remaining $k'$ biclusters $\mathcal{B} = (\mathcal{R}_1, \mathcal{C}_1), \ldots,  (\mathcal{R}_{k'}, \mathcal{C}_{k'})$ where the rows of $\mathcal{S}$ in each bicluster $\alpha_i$ correspond to $\mathcal{R}_i = \{j \in \alpha_i \mid j \in Z_[1:r]\}$, and the columns/features of $\mathcal{G}$ selected are $\mathcal{C}_i = \{ j \in \alpha_i \mid j \in Z_[r+1:r+c] \}$
\end{algorithmic}
\end{algorithm}

We augment this strategy to jointly perform biclustering while automatically determining the number of clusters. The details are outlined in \autoref{algo:bicluster}, where $\mathcal{S}$ are the sample inputs and $\mathcal{G}$ the set of input features that make the rows and columns, respectively, of the matrix $A$. Lines 1 through 7 proceed in the same manner as standard SpectralCoClustering, except we use a larger set of features for the matrix $Z$. We then use a Variational Gaussian Mixture Model (VGMM) \cite{Rasmussen2000,export:67239} to perform the clustering instead of $k$-means, as the VGMM algorithm can automatically determine the number of clusters to use. In particular, we use a diagonal covariance constraint on the GMM so that if we have extraneous clusters, the VGMM can learn to ignore the extra dimensions in $Z$, which should exhibit homogeneity in the coefficients due to the excessive number of dimensions no longer forming a meaningful cut in the bipartite graph clustering. 

On line 11 we use the probabilities of cluster membership computed by the VGMM to select any row (sample) with a probability $\geq 1/(k+1)$ of belonging to each bicluster. We use $k+1$ in the denominator so that multi-bicluster membership can still occur with $k=2$ biclusters.  

In lines 12-18 of the algorithm, we perform removal of extraneous clusters. First we remove clusters that contain only rows from $A$\textbackslash$\mathcal{S}$, and only columns from $A$\textbackslash$\mathcal{G}$, as these biclusters are degenerate and uninformative. In the next step we filter out clusters that contain less than 5 rows or columns from $A$ as being spurious, unless the largest clusters contain fewer than 5 rows/columns, at which point we set the limit to the largest observed. This allows us to avoid spurious clusters while adapting to work in scenarios with small sample sizes that occur in malware analysis (e.g., $\leq 10$ samples), but would not normally be of interest in standard biclustering applications. 

\section{AutoYara Design}\label{sec:autoyara}

In designing the AutoYara tool, a number of design constraints informed our approach. First, the tool must be light-weight enough that it can run on low resource machines (e.g., a laptop with 4 GB of RAM or less) to support the maximal number of analysts, who do not always have significant compute resources available. Toward this goal, we needed to minimize memory use and model size in memory, as well as reliance on any GPU resources. This allows analysts who take "fly-away" kits with them to unfamiliar networks to begin investigations into the network, encountering whatever novel malware that may be present \cite{Nguyen2019_filename_malicious}. We also need the tool to produce Yara rules within minutes, as our experience has been that analysts will not, in general, use tools requiring them to wait hours or more. Finally, we need to produce Yara rules that can be interpreted by analysts. This makes it is possible for analysts to gain insights that may aid their work by inspecting the rules or even modify the rules to improve them. 

We assume the user will provide multiple files that share some intrinsic nature (e.g., same malware family), which we would like to identify with a Yara rule. We focus on building Yara rules based on specific byte patterns. 
For this reason we use large byte $n$-grams, where $n\in [8, 1024]$. Prior work has developed an algorithm to extract the top-$k$ most frequent $n$-grams for large values of $n$ with limited memory, in time almost invariant to $n$, and has noted that these large byte-grams are interpretable to malware analysts \cite{Kilograms_2019}.

To make sure that we only consider interesting $n$-grams for rule construction, we will perform filtering of the selected $n$-grams. First, we will use a large training corpus of 600,000 files to find generally frequent $n$-grams. If an $n$-gram is frequent across a large portion of these files, it is unlikely to make a good signature --- as signatures need to be specific. To store these compactly, we will use a  Bloom Filter for each $n$-gram size, storing the top 1 million most frequent $n$-grams for $n \in 8, 16, 32, \ldots, 1024$ if they occur in more than 0.1\% of the training files.

\begin{algorithm}[!ht]
\caption{Filter Simple } \label{algo:filter}
\begin{algorithmic}[1]
\Require  Set of files/data points $\mathcal{S}$, and set of n-gram features $\mathcal{G}$, bloom filters $F_{n}$
\ForAll{byte n-grams $g_i \in \mathcal{G}$}
    \State $z = \sum_{j \in g_i} \mathbbm{1}[j = 0x00] + \mathbbm{1}[j = 0xFF]$ \Comment{Count number of bytes equal to 0 or 255}
    \If{$z \geq |g_i|/2$}
        \State remove $g_i$
    \ElsIf{$H(g_i) \leq 1.0$} \Comment{Byte entropy of too small}
        \State remove $g_i$
    \ElsIf{$g_i \in$ bloom filter $F_{|g_i|}$}
        \State remove $g_i$
    \EndIf
\EndFor
\State For each pair of n-grams $g_i$ and $g_j$, if they both occur in exactly the same files in $\mathcal{S}$, keep the n-gram with the highest entropy and discard the other. 
\State \textbf{return} remaining n-grams $g_i$ that were not removed. 
\end{algorithmic}
\end{algorithm}

We use two other strategies for removing $n$-grams unlikely to be useful for clustering. First we consider the entropy of an $n$-gram $x$, as given in \autoref{eq:entropy}, where $P_i(x)$ denotes the proportion of bytes with value $i$ (i.e, $P_i(x) = \sum_{j=0}^{n-1} \mathbbm{1}[x[j] = i]/n$). 

\begin{equation} \label{eq:entropy}
    H(x) = - \sum_{i=0}^{255} P_i(x) \cdot \log \left( P_i(x) \right)
\end{equation}

The byte entropy of a sequence would then be in the range of [0,8], with 8 corresponding to content that appears completely random, and 0 for the same value repeated alone.  For context, natural language text usually has an entropy value $\approx 4$. We use a filter of 1.0 to remove n-grams. We also check if more than half of the bytes have the value 0 or 0xFF, which are commonly used in padding and can be unreliable. This gives us a simple filtering strategy given by \autoref{algo:filter}. 

\begin{algorithm}[!ht]
\caption{AutoYara} \label{algo:autoyara}
\begin{algorithmic}[1]
\Require Benign \& malicious training corpora $\mathcal{C}$. Top-$k$ value $k$. Initial n-gram size $n_{\text{l}}$, minimum entropy $m_{h}$, and filter threshold $f_t$. 
\Function{BuildIndex}{Corpus $\mathcal{C}$}
    \For{$i \in  8, 16, 32, 64, 128, \ldots, 1024$}
        \State  $\mathcal{G} \gets$ Find top-$k$ most frequent $n_{\text{i}}$-grams using Kilo-Gram Algorithm\cite{Kilograms_2019}. 
        \State Create new counting Bloom Filter $F_{n_i}$
        \ForAll{$g \in \mathcal{G}$ } \Comment{Populate Bloom Filter}
            \State $c \gets$\Call{Count}{$g$}
            \State $F_{n_i}$.\Call{Insert}{$g$, $c$}
        \EndFor
    \EndFor
\EndFunction
\Function{BuildYaraRule}{Sample files $\mathcal{S}$}
    \State Current Best Rule $R_{best} \gets 0$
    \State Current best score $s_{best} \gets 0$
    
    \For{$i \in 8, 16, 32, 64, 128, \ldots, 1024$}
        \State  $\mathcal{G} \gets$ Find top-$k$ most frequent $n_{\text{i}}$-grams using \cite{Kilograms_2019}. 
        \State \Call{FilterSimple}{$\mathcal{G}$, $F_{n_i}$ $m_h$, $f_t$} \Comment{Using \autoref{algo:filter}}
        \State $\mathcal{B} \gets $ \Call{Bicluster}{$\mathcal{S}$, $\mathcal{G}$}  \Comment{Using \autoref{algo:bicluster}}
        \State Create empty rule $R \gets \emptyset$ 
        \State Create set $covered \gets \emptyset$
        \For{Row-column tuple $\mathcal{R}, \mathcal{C} \in \mathcal{B}$}
            \State Let \Call{Count}{$c$} be the number of times feature $c \in \mathcal{C}$ be the number of files in $\mathcal{S}$ that  feature/$n_i$-gram $c$ occurred in.
            \State Let $\sigma_{i:j}$ indciate the variance of \Call{Count}{$c_i$}, \Call{Count}{$c_{i+1}$}, $\ldots$, \Call{Count}{$c_j$} where the counts are sorted from minum to maximum. 
            \State $t \gets \underset{s}{\arg \min } s \cdot \sigma_{1: s}^{2}+(n-s) \cdot \sigma_{s: n}^{2}$
            \State $R \gets R \land (t \text{ of } \mathcal{C})$ 
            \State $covered \gets covered \cup \mathcal{R}$
        \EndFor
        \State $s \gets \frac{|covered|}{|\mathcal{S}|} \cdot \frac{\min \left(5, \left| \bigcup_{\mathcal{R}, \mathcal{C} \in \mathcal{B}} \bigcup_{\forall c \in \mathcal{C}} \right| \right)}{5}$
        \If{$s > s_{best}$} \Comment{We found a better YaraRule}
            \State $s_{best} \gets s$
            \State $R_{best} \gets R$
        \EndIf
    \EndFor
    \State \Return Yara rule $R_{best}$
\EndFunction
\end{algorithmic}
\end{algorithm}

We now have all the information we need to specify our new AutoYara algorithm for constructing Yara rules from raw bytes. This algorithm is shown in \autoref{algo:autoyara}. First the \textit{BuildIndex} function creates Bloom Filters for each value of $n$ considered. These take up about 33 MB of disk each, and is done once in advance. 

The \textit{BuildYaraRule} does the majority of the work to create a Yara rule that hopefully matches the set of files given in $\mathcal{S}$. For every $n$-gram size $n$, we will extract the top-$k$ most frequent grams, use \autoref{algo:filter} to remove unlikely features, and then \autoref{algo:bicluster} to create a biclustering of the data. As described in \autoref{sec:bicluster}, each feature used within a bicluster is merged into a larger clause by "and"ing the terms together, and "or"ing the biclusters together. 

To improve the quality of the biclustering, on lines 19-21, we do not naively "and" every $n$-gram found within a bicluster. Instead, we select a threshold $t$ of the rules to be found, as not every feature will always appear in a new file. This threshold is selected based on the same heuristic used in decision trees. We take the number of occurrences of each feature $c \in \mathcal{C}$ in the input samples and sort them from fewest to most frequent occurrences. We then find the split that minimizes the variance in counts and use that as the threshold for the number of features/$n$-grams used to satisfy this specific clause. This approach assumes that there will be some set of n-grams that are useful features and common, and a second population of n-grams that are excessively frequent because they are generically frequent. 

These biclusters are evaluated based on the coverage of the input files $\mathcal{S}$, and we rely on the length of the rules and the n-grams themselves to avoid false positives. If a developed Yara rule $R$ obtains 100\% coverage, it will be selected as the final rule. We note that in line 20, we also include a penalty based on the number of $n$-gram components used in the candidate rule. If $|R|$ denotes the number of $n$-gram features used in a rule, we are penalizing the rule by a factor of $|R|/5$ if $|R| < 5$. This is done to avoid false positives, as we prefer rules with more terms to bias our method to low false-positive rates. 

Our final implementation of AutoYara is in Java to enable use on multiple operating systems and fast execution time. It can be found at \url{https://github.com/NeuromorphicComputationResearchProgram}, and uses the JSAT library \cite{JMLR:v18:16-131} to implement the biclustering.

\section{Datasets} \label{sec:datasets}

To train our AutoYara system, we will use the Ember 2017 corpus \cite{Anderson2018}, which contains a training set of 300,000 benign and 300,000 malicious files. We use only this data to create our Bloom Filters used during the entire process. By using both benign and malicious files to build our Bloom Filters, we obtain better coverage to hopefully ensure a low false positive rate. We explicitly do not use the test set of Ember at any point. First, the test set is organized by benign vs. malicious, which is not the ultimate goal of AutoYara. More importantly, we want to maximize the difficulty of our evaluation to better judge the generalization properties of AutoYara. By using different data that was collected from different sources, we decrease the similarity between training and testing data, thus getting a better judgment of generalization and utility to real life situations in which analysts would use this tool \cite{Rossow2012,Allix:2016:EAM:2887211.2887272,Maiorca2018,raff2020survey}.  

We expected AutoYara to be run only on malicious files that are related in some manner (e.g., a malware family) of interest to the analyst. For this purpose, we construct a larger dataset using VirusShare \cite{VirusShare} to get a large corpus of malware. To determine the family labels for this corpus, we use the AVClass tool \cite{Sebastian2016} to coalesce the outputs of multiple anti-virus products from VirusTotal \cite{Virustotal} reports collected by \cite{Seymour2016}.
This gives us a larger dataset to perform evaluations on, but is still susceptible to noisy labels provided by AV products, and imperfect coalescing by AVClass in selecting a final label. As such this is our noisiest test data. 

Because we want to investigate the impact on the number of samples with respect to tool accuracy, we decided to select malware families from VirusShare that had at least 10,000 examples each. This gave us 184 total malware families. For each family, we randomly selected 2,000 files to be a test set and sample from the remainder for training. For each family, we will ask AutoYara (trained on Ember) to create a Yara rule, for which we will determine the true-positive rate from the 2,000 samples. 

To measure false-positive rates, we will evaluate each generated rule on the remaining 183 other families. However, it is possible that the false positive rate may differ significantly between benign and malicious out-of-class samples. To maximize the difficulty and best judge generalization, we use the 200,000 benign and 200,000 malicious files used by \cite{raff_shwel}, as a special held-out corpus. 
This allows us to measure false-positive rates as low as $10^{-6}$, and we will routinely see that our method can obtain exactly 0 false positives. 

We will also use a dataset provided by 
Elastic 
taken from a production environment. In particular, 
Elastic 
had 24 different malware families for which there was a production need to write Yara rules to identify these specific families. Two expert analysts (A and B) with $\geq 5$ years experience recorded their time/progress on these families through the course of a normal workday, allowing us to show how AutoYara can save an analyst over 44\% of their time in rule construction to better meet their mission requirements. A third analyst (C) with $\geq 2$ years experience was given no other tasking but to process all 24 families and create rules for them. Analyst C had not previously used Yara, but had prior reverse engineering experience.

\section{Experimental Validation} \label{sec:results}

Now that we have described our AutoYara concept, and reviewed the procedure for determining the search policy AutoYara uses to build Yara rules, we will investigate its performance on three tasks. For these tasks, the guidance 
Elastic
uses is that a rule needs to have a false positive (FP) rate $\leq 0.1\%$ for a rule to be \textit{potentially} useful. The utility of a rule depends on the true positive (TP) rate and the degree of need for the rule. To simplify analysis, we will describe the performance of the rules as a whole using the $F_\beta$ metric, as given in \autoref{eq:f_beta}. The $F_\beta$ score gives us a measure where we can state that a true positive is more/less important than a false positive. As such, we use $\beta=0.001$, corresponding to each false positive being worse than a false negative in line with our desire that FP should be $\leq 0.1\%$. 

\begin{equation} \label{eq:f_beta}
    F_{\beta}=\frac{\left(1+\beta^{2}\right) \cdot \text { true positive }}{\left(1+\beta^{2}\right) \cdot \text { true positive }+\beta^{2} \cdot \text { false negative }+\text { false positive }}
\end{equation}

First, we will compare AutoYara with  YarGen and the Greedy approach of prior works on the 184 families that were discussed above. This will show that AutoYara dominates YarGen with respect to $F_\beta$ score, and that the Greedy approach does not work when only a small number of samples are given. Second, we will do a larger scale test that simulates behavior in a malware hunt situation, where an analyst deploys to a remote network and has the goal of finding malware on a network that is known or suspected to be compromised. This will show that we can generate rules with extremely low false positives, even when querying against $\geq$90,000,000 files. Third, we will perform a true real-world task with AutoYara compared to two professional analysts performing their work at 
Elastic,
which shows that AutoYara can be a useful tool as it matches professional analyst performance on a number of malware families. 

\subsection{Large Scale Testing}

In this first experiment, we are interested in the applicability of all methods to creating Yara rules over a large range of families and sample sizes ranging from $n=2$ to $n=2^{12}$ examples. The Greedy approach quickly becomes disqualified due to not generating enough viable rules, producing $\leq 17$ samples with a FPR$\leq 0.1\%$ for $n\geq 8$ samples. VxSig is also disqualified from this section because it is too computational demanding to run, requiring an estimated \textit{603 years} to run over all settings. This leaves YarGen and AutoYara which can be tested across all 184 families and sample sizes. AutoYara produced 50-114 viable rules for each value of $n$, and YarGen produced $6-121$ variable rules for each value of $n$. 

\begin{figure}[!ht]
    \centering
    \adjustbox{max width=\columnwidth}{%
    \includegraphics[]{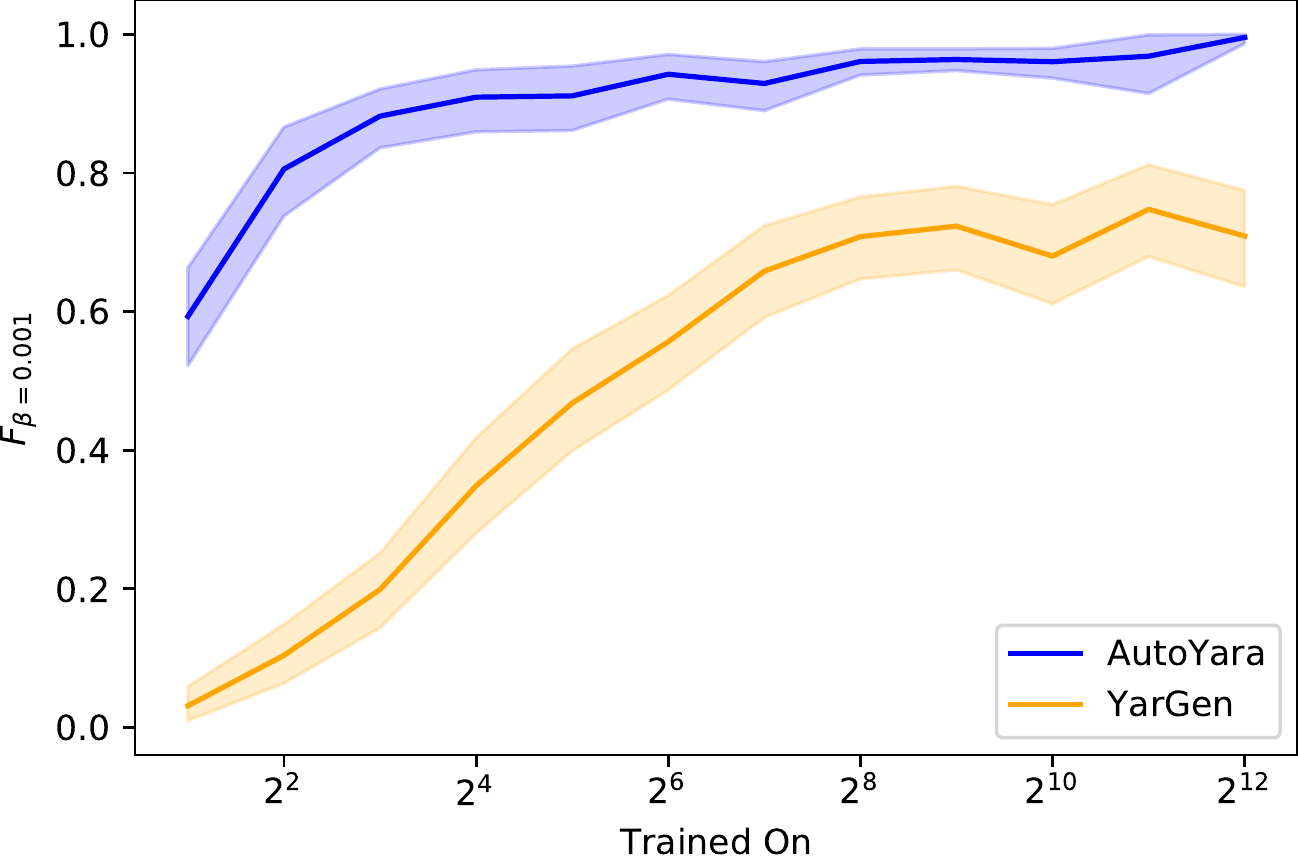}
    }
    \caption{Comparison of AutoYara and YarGen average $F_\beta$ score on 184 malware families. Confidence intervals are constructed using the Jacknife resampling. }
    \label{fig:yargen_vs_autoyara}
\end{figure}

 We compare the average $F_\beta$ score between AutoYara and YarGen over the viable rules ($\leq 0.1\%$ FPR) generated from VirusShare in \autoref{fig:yargen_vs_autoyara}. This shows that AutoYara produces rules of a significantly higher quality than YarGen. In fact, rules generated by YarGen are not generally usable until 128 samples are given, which is more than an analyst would have available in most situations. This is particularly true for hunt missions, which we will discuss in Section~\ref{sec:retro_hunt}. 

\begin{figure}[!ht]
    \centering
    \adjustbox{max width=\columnwidth}{%
    \includegraphics[]{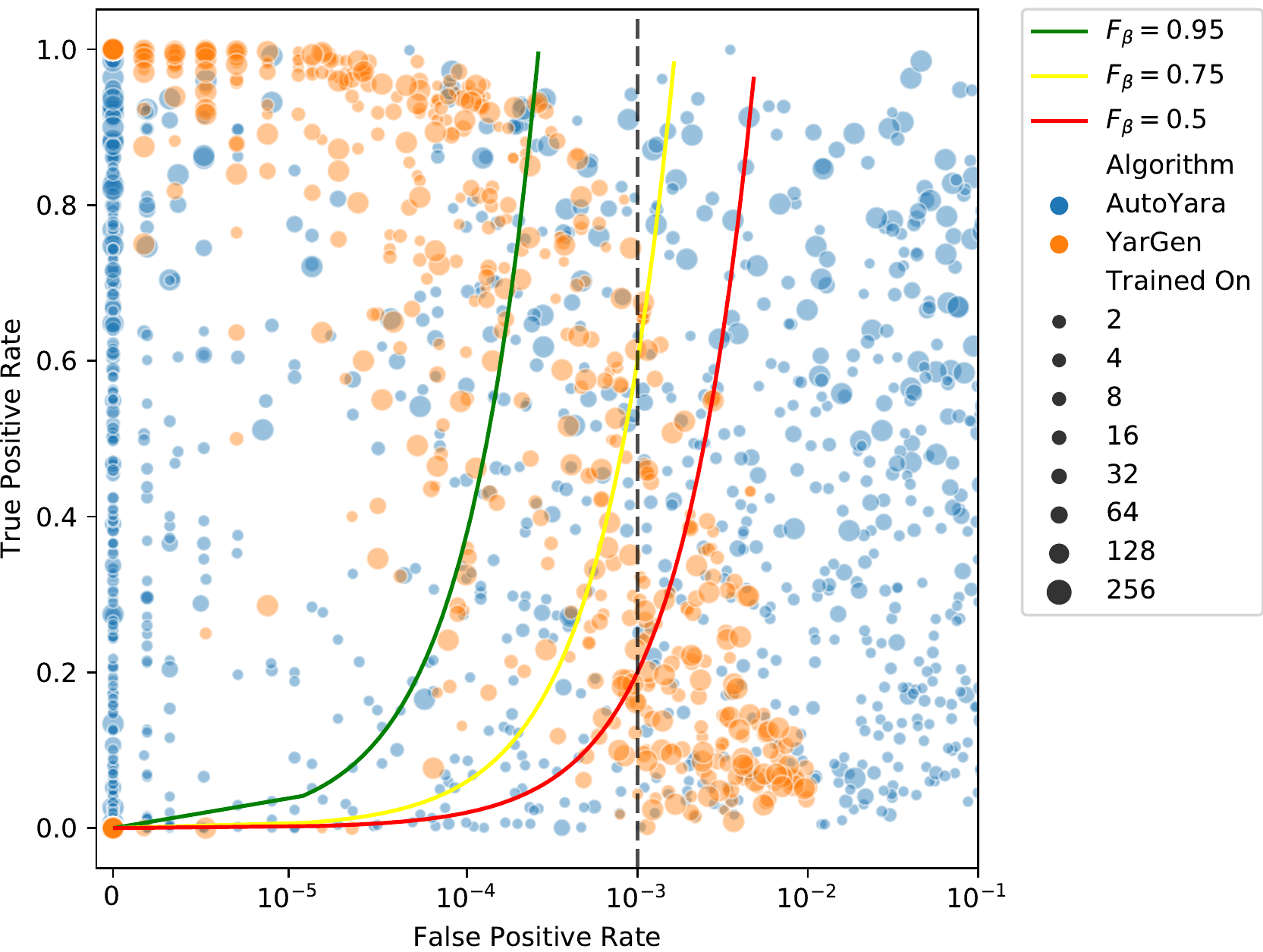}
    }
    \caption{Scatter plot of the $F_\beta$ score achieved by individual rules created from  AutoYara and YarGen on 184 malware families. Dot size indicates the number of training samples used to create the rule. The black dashed line shows the desired minimum false positive rate of 0.1\%, and the solid lines 
    show the curve to achieve a minimum $F_\beta = $ \textcolor{OliveGreen}{0.95}, \textcolor{Goldenrod}{0.75}, \textcolor{red}{0.5}.  
    }
    \label{fig:yargen_vs_autoyara_detailed}
\end{figure}

To further explain why AutoYara produces a higher score, and that false positives alone are insufficient, we plot the $F_\beta$ score of all generated rules in \autoref{fig:yargen_vs_autoyara_detailed}, where the x-axis shows the false positive rate (log scale) and the y-axis the true positive rate (linear scale). The size of the dots show how many files were used to create the rule, ranging from 2 to 256 for visual clarity. YarGen contains a a majority of rules all located at the bottom left corner with 0\% FP and 0\% TP, making the rules ineffective. The only rules that obtain higher TP rates are those trained on more files, and we can see that YarGen suffers a bias of decreasing TP rate as the FP rate increases. This is the worst case performance behavior and shows how it biases YarGen to obtaining lower $F_\beta$ scores. 

In contrast, AutoYara produces several rules with 0 false positives at a variety of true positive rates. It can also be seen that as the TP rate increases at 0 FPs, the number of samples trained on increases, causing the average $F_\beta$ score to increase with sample size. This demonstrates our strategy is succeeding in obtaining low FPs over a wide range of sample sizes. In the cases where AutoYara does not produce as good a rule in terms of FPs, it is not biased toward also lowering its TP rate --- allowing it to obtain generally better rules for any fixed desired FP rate.

Due to YarGen's performance, and the high monetary and human time cost to perform the experiments detailed in the following sections, we do not further consider it in comparison to AutoYara. 

\subsection{Retro Hunt Results} \label{sec:retro_hunt}

Our second round of testing is motivated by a common type of "hunt" team operation, where analysts will deploy to networks that they are not familiar with to search for malware present on those networks. Such an operation may be spurred by knowledge or supposition of a compromised network, and may or may not include knowledge of what kind of malware is being searched for. 

\begin{table}[!ht]
\caption{Results analyzing files returned by VirusTotal Retro Hunt given a Yara rule generated using AutoYara and VxSig. Best method shown in \textbf{bold}. TP\%, FP\% on malware, and FP\% on benign ware is estimated based on up to 50 samples returned by VT.  } \label{tbl:retro_hunt}
\adjustbox{max width=\columnwidth}{%
\begin{tabular}{@{}clrrrr@{}}
\toprule
Family/Samples                        & \multicolumn{1}{c}{Method} & \multicolumn{1}{c}{New VT Hits} & \multicolumn{1}{c}{TP\%} & \multicolumn{1}{c}{FP\% Mal} & \multicolumn{1}{c}{FP\% Benign} \\ \midrule
\multicolumn{1}{l}{APT 26 Fancy Bear} & AutoYara                   & $\geq$10,000                    & 0                        & 92                           & 8                               \\
28                                    & \textbf{VxSig}            & \textbf{4,079}                  & \textbf{0}               & \textbf{98}                  & \textbf{2}                      \\ \midrule
\multicolumn{1}{l}{APT 28}            & \textbf{AutoYara}          & \textbf{$\geq$10,000}           & \textbf{2}               & \textbf{54}                  & \textbf{44}                     \\
61                                    & VxSig                     & N/A                             & ---                      & ---                          & ---                             \\ \midrule
\multicolumn{1}{l}{ATMDtrack\_DPRK}   & \textbf{AutoYara}          & \textbf{2}                      & \textbf{100}             & \textbf{0}                   & \textbf{0}                      \\
3                                     & VxSig                     & 0                               & ---                      & ---                          & ---                             \\ \midrule
\multicolumn{1}{l}{CloudHopper/APT10} & \textbf{AutoYara}          & \textbf{5,118}                  & \textbf{0}               & \textbf{100}                 & \textbf{0}                      \\
229                                   & VxSig                     & 1,251                           & 0                        & 92                           & 6                               \\ \midrule
\multicolumn{1}{l}{CobaltGroup}       & AutoYara                   & $\geq$10,000                    & 0                        & 98                           & 2                               \\
9                                     & \textbf{VxSig}            & \textbf{1}                      & \textbf{0}               & \textbf{100}                 & \textbf{0}                      \\ \midrule
\multicolumn{1}{l}{Dridex}            & \textbf{AutoYara}          & \textbf{36}                     & \textbf{100}             & \textbf{0}                   & \textbf{0}                      \\
5                                     & VxSig                     & 1                               & 100                      & 0                            & 0                               \\ \midrule
\multicolumn{1}{l}{Dyre}              & \textbf{AutoYara}          & \textbf{1}                      & \textbf{100}             & \textbf{0}                   & \textbf{0}                      \\
8                                     & VxSig                     & $\geq$10,000                    & 2                        & 92                           & 16                              \\ \midrule
\multicolumn{1}{l}{EquationGroup}     & \textbf{AutoYara}          & \textbf{4}                      & \textbf{100}             & \textbf{0}                   & \textbf{0}                      \\
10                                    & VxSig                     & 1                               & 100                      & 0                            & 0                               \\ \midrule
\multicolumn{1}{l}{GamaredonGroup}    & \textbf{AutoYara}          & \textbf{26}                     & \textbf{100}             & \textbf{0}                   & \textbf{0}                      \\
7                                     & VxSig                     & $\geq$10,000                    & 0                        & 76                           & 24                              \\ \midrule
\multicolumn{1}{l}{GrandCrab}         & \textbf{AutoYara}          & \textbf{62}                     & \textbf{100}             & \textbf{0}                   & \textbf{0}                      \\
7                                     & VxSig                     & N/A                             & ---                      & ---                          & ---                             \\ \midrule
\multicolumn{1}{l}{GreenBugAPT}       & AutoYara                   & 5                               & 100                      & 0                            & 0                               \\
4                                     & VxSig                     & 5                               & 100                      & 0                            & 0                               \\ \midrule
\multicolumn{1}{l}{GreyEnergy}        & AutoYara                   & 2                               & 100                      & 0                            & 0                               \\
3                                     & VxSig                     & 2                               & 100                      & 0                            & 0                               \\ \midrule
\multicolumn{1}{l}{OlympicDestroyer}  & \textbf{AutoYara}          & \textbf{2}                      & \textbf{100}             & \textbf{0}                   & \textbf{0}                      \\
4                                     & VxSig                     & 5                               & 80                       & 20                           & 0                               \\ \midrule
\multicolumn{1}{l}{Shamoon}           & AutoYara                   & 0                               & ---                      & ---                          & ---                             \\
2                                     & \textbf{VxSig}            & \textbf{1}                      & \textbf{100}             & \textbf{0}                   & \textbf{0}                      \\ \midrule
\multicolumn{1}{l}{Sofacy}            & \textbf{AutoYara}          & \textbf{453}                    & \textbf{6}               & \textbf{54}                  & \textbf{40}                     \\
3                                     & VxSig                     & $\geq$10,000                    & 0                        & 72                           & 28                              \\ \midrule
\multicolumn{1}{l}{Sugar}             & \textbf{AutoYara}          & \textbf{923}                    & \textbf{100}             & \textbf{0}                   & \textbf{0}                      \\
17                                    & VxSig                     & 1,764                           & 56                       & 44                           & 0                               \\ \midrule
\multicolumn{1}{l}{Thrip}             & AutoYara                   & $\geq$10,000                    & 0                        & 98                           & 2                               \\
76                                    & VxSig                     & $\geq$10,000                    & 0                        & 98                           & 2                               \\ \midrule
\multicolumn{1}{l}{Turla (Uroburos)}  & \textbf{AutoYara}          & \textbf{$\geq$10,000}           & \textbf{0}               & \textbf{90}                  & \textbf{10}                     \\
11                                    & VxSig                     & $\geq$10,000                    & 0                        & 88                           & 12                              \\ \midrule
\multicolumn{1}{l}{WannaCry}          & \textbf{AutoYara}          & \textbf{6520}                   & \textbf{100}             & \textbf{0}                   & \textbf{0}                      \\
2                                     & VxSig                     & N/A                             & ---                      & ---                          & ---                             \\ \midrule
\multicolumn{1}{l}{Petya}             & AutoYara                   & $\geq$10,000                    & 0                        & 2                            & 98                              \\
5                                     & \textbf{VxSig}            & \textbf{482}                    & \textbf{2}               & \textbf{98}                  & \textbf{0}                      \\ \bottomrule
\end{tabular}
}
\end{table}

When malware is found on the network, analysts often begin to search the rest of the network for malware of the same type. AutoYara can be used to accelerate this task by constructing rules from a small set of observed files, and then analysts can use Yara with existing tooling to scan the larger network. In this application false positives on other malicious families are still useful, though not the target. Any malware found on the network is of interest to analysts and may be important, even if it was not of the same type that was expected. Only benign false positives are a nuisance in this case, as they divert analyst time into investigating non-issues. 

\begin{table*}
\caption{Comparing 
against three professional analysts. For each cell, we report the family coverage rate (top), false positive rate (middle) and time required in minutes to create the rule (bottom). Analyst B used YarGen as part of their work. 
Highlighted columns indicate an automated tool produced a usable rule ($\leq 0.1\%$ FP).  \textbf{Bold} indicates best results for tooling. 
} \label{tbl:bot_v_human}
\adjustbox{max width=1.0\textwidth}{%
\begin{tabular}{@{}lcccccccccccccccccccccccc@{}}
\toprule
                           & \mcrot{1}{l}{60}{baldr} & \mcrot{1}{l}{60}{baofa} & \mcrot{1}{l}{60}{bkff} & \mcrot{1}{l}{60}{conju} & \mcrot{1}{l}{60}{darkvnc} & \mcrot{1}{l}{60}{dragonmess} & \mcrot{1}{l}{60}{ertfor} & \mcrot{1}{l}{60}{firefly} & \mcrot{1}{l}{60}{jongiti} & \mcrot{1}{l}{60}{ladyoffice} & \mcrot{1}{l}{60}{navattle} & \mcrot{1}{l}{60}{nezchi} & \mcrot{1}{l}{60}{olympic destroyer} & \mcrot{1}{l}{60}{phds} & \mcrot{1}{l}{60}{phtominer} & \mcrot{1}{l}{60}{pikachu} & \mcrot{1}{l}{60}{plurox} & \mcrot{1}{l}{60}{potukorp} & \mcrot{1}{l}{60}{sekur} & \mcrot{1}{l}{60}{subroate} & \mcrot{1}{l}{60}{wininf} & \mcrot{1}{l}{60}{wuca} & \mcrot{1}{l}{60}{xpantispyware} & \mcrot{1}{l}{60}{zcash} \\
\midrule
                            &                                        & 85.71                                  & 90.01                                & 100                                    &         & 84.62                                  & 15.38                                   & 80                                       & 88.24                                  & 100                                   & 84.62                                    & 100                         & 75.00                                  & 88.24                                  & 100                                  & 83.33                                  &        & 81.25                                   &                                        & 46.67                                  &                                        &                                        &                                        & 100                                  \\
                            &                                        & 0                                      & 0                                    & 0                                      &         & 0                                      & 0                                       & 0                                        & 0                                      & 0                                     & 0                                        & 0                           & 0                                      & 0                                      & 0                                    & 0                                      &        & 0                                       &                                        & 0                                      &                                        &                                        &                                        & 0                                    \\
\multirow{-3}{*}{Analyst A} &                                        & 8                                      & 9                                    & 10                                     &         & 10                                     & 8                                       & 4                                        & 6                                      & 8                                     & 9                                        & 10                          & 30                                     & 14                                     & 5                                    & 9                                      &        & 10                                      &                                        & 10                                     &                                        &                                        &                                        & 10                                   \\ \cmidrule(l){2-25} 
                            & 100                                    &                                        & 90.01                                &                                        & 50      &                                        &                                         & 100                                      &                                        & 100                                   &                                          & 100                         & 62.50                                  &                                        & 100                                  & 100                                    & 0      &                                         &                                        &                                        &                                        &                                        &                                        & 100                                  \\
                            & 11.79                                  &                                        & 0                                    &                                        & 13.96   &                                        &                                         & 2.16                                     &                                        & 0.05                                  &                                          & 0                           & 0                                      &                                        & 0                                    & 0                                      & 0.025  &                                         &                                        &                                        &                                        &                                        &                                        & 0.025                                \\
\multirow{-3}{*}{Analyst B} & 78                                     &                                        & 5                                    &                                        & 70      &                                        &                                         & 40                                       &                                        & 20                                    &                                          & 20                          & 15                                     &                                        & 5                                    & 10                                     & 30     &                                         &                                        &                                        &                                        &                                        &                                        & 25                                   \\ \cmidrule(l){2-25} 
                            & 66.67                                  & 64.29                                  & 63.64                                & 100                                    & 0       & 84.62                                  & 0                                       & 40.00                                    & 88.24                                  & 100                                   & 61.54                                    & 100                         & 75                                     & 88.24                                  & 100                                  & 83.33                                  & 0      & 81.25                                   & 20                                     & 26.67                                  & 100                                    & 16.67                                  & 61.54                                  & 100                                  \\
                            & 0.1098                                 & 0.00025                                & 0.014                                & 0                                      & 0.004   & 0.003                                  & 0                                       & 0                                        & 0                                      & 0                                     & 0                                        & 0.00125                     & 0                                      & 0.001                                  & 0                                    & 0                                      & 0      & 0.00175                                 & 0.16425                                & 3.0745                                 & 0.019                                  & 0.00325                                & 0                                      & 0                                    \\
\multirow{-3}{*}{Analyst C} & 25                                     & 15                                     & 20                                   & 20                                     & 35      & 20                                     & 25                                      & 40                                       & 25                                     & 20                                    & 20                                       & 25                          & 20                                     & 15                                     & 25                                   & 20                                     & 25     & 20                                      & 40                                     & 60                                     & 10                                     & 60                                     & 40                                     & 20                                   \\ \midrule
                            & \cellcolor[HTML]{C0C0C0}\textbf{66.67} & 64.26                                  & 0                                    & \cellcolor[HTML]{C0C0C0}23.53          & 50      & 0                                      & 0                                       & 0                                        & \cellcolor[HTML]{C0C0C0}\textbf{52.94} & \cellcolor[HTML]{C0C0C0}\textbf{100}  & \cellcolor[HTML]{C0C0C0}23.08            & 0                           & \cellcolor[HTML]{C0C0C0}75             & 0                                      & \cellcolor[HTML]{C0C0C0}100          & \cellcolor[HTML]{C0C0C0}\textbf{83.33} & 0      & \cellcolor[HTML]{C0C0C0}\textbf{87.5}   & 0                                      & 26.67                                  & \cellcolor[HTML]{C0C0C0}100            & 0                                      & 0                                      & 0                                    \\
                            & \cellcolor[HTML]{C0C0C0}\textbf{0}     & 0.2243                                 & 0                                    & \cellcolor[HTML]{C0C0C0}0              & 35.65   & 0                                      & 0                                       & 0                                        & \cellcolor[HTML]{C0C0C0}\textbf{0}     & \cellcolor[HTML]{C0C0C0}\textbf{0}    & \cellcolor[HTML]{C0C0C0}0                & 16.65                       & \cellcolor[HTML]{C0C0C0}0              & 0                                      & \cellcolor[HTML]{C0C0C0}0            & \cellcolor[HTML]{C0C0C0}\textbf{0}     & 0      & \cellcolor[HTML]{C0C0C0}\textbf{0.0005} & 0                                      & 0.4475                                 & \cellcolor[HTML]{C0C0C0}0.0755         & 0                                      & 0                                      & 0                                    \\
\multirow{-3}{*}{VxSig}    & \cellcolor[HTML]{C0C0C0}\textbf{10.2}  & 1386.4                                 & 26.2                                 & \cellcolor[HTML]{C0C0C0}474.1          & 24.2    & 35.9                                   & 32.1                                    & 24                                       & \cellcolor[HTML]{C0C0C0}\textbf{53.55} & \cellcolor[HTML]{C0C0C0}\textbf{11.9} & \cellcolor[HTML]{C0C0C0}57.7             & 17.6                        & \cellcolor[HTML]{C0C0C0}34.7           & 47.8                                   & \cellcolor[HTML]{C0C0C0}7.5          & \cellcolor[HTML]{C0C0C0}\textbf{20.1}  & 6.0    & \cellcolor[HTML]{C0C0C0}\textbf{159.4}  & 20.8                                   & 46.1                                   & \cellcolor[HTML]{C0C0C0}68.2           & 28.0                                   & 42.1                                   & 4.1                                  \\ \cmidrule(l){2-25} 
                            & 100                                    & 100                                    & \cellcolor[HTML]{C0C0C0}90.01        & 100                                    & 100     & 100                                    & 92.31                                   & 80                                       & \cellcolor[HTML]{C0C0C0}88.24          & 100                                   & 84.62                                    & 100                         & 100                                    & 100                                    & \cellcolor[HTML]{C0C0C0}\textbf{100} & 100                                    & 100    & 93.75                                   & 100                                    & 100                                    & 100                                    & 100                                    & 100                                    & \cellcolor[HTML]{C0C0C0}\textbf{100} \\
                            & 83.69                                  & 10.34                                  & \cellcolor[HTML]{C0C0C0}0            & 83.25                                  & 83.71   & 6.84                                   & 81.09                                   & 13.11                                    & \cellcolor[HTML]{C0C0C0}0              & 83.72                                 & 16.45                                    & 2.025                       & 10.62                                  & 12.80                                  & \cellcolor[HTML]{C0C0C0}\textbf{0}   & 0.725                                  & 83.74  & 83.36                                   & 70.95                                  & 83.40                                  & 83.40                                  & 83.40                                  & 76.24                                  & \cellcolor[HTML]{C0C0C0}\textbf{0}   \\
\multirow{-3}{*}{Greedy}    & 0.5                                    & 0.7                                    & \cellcolor[HTML]{C0C0C0}0.9          & 1.4                                    & 0.6     & 1.2                                    & 0.3                                     & 0.3                                      & \cellcolor[HTML]{C0C0C0}0.4            & 0.5                                   & 0.6                                      & 1.2                         & 1.9                                    & 1.0                                    & \cellcolor[HTML]{C0C0C0}\textbf{0.9} & 0.4                                    & 0.5    & 2.2                                     & 0.4                                    & 0.2                                    & 0.2                                    & 0.3                                    & 0.5                                    & \cellcolor[HTML]{C0C0C0}\textbf{1.4} \\ \cmidrule(l){2-25} 
                            & 66.67                                  & \cellcolor[HTML]{C0C0C0}\textbf{14.29} & \cellcolor[HTML]{C0C0C0}\textbf{100} & \cellcolor[HTML]{C0C0C0}\textbf{94.12} & 50.00   & \cellcolor[HTML]{C0C0C0}\textbf{76.92} & \cellcolor[HTML]{C0C0C0}\textbf{23.07}  & \cellcolor[HTML]{C0C0C0}\textbf{20}      & \cellcolor[HTML]{C0C0C0}41.18          & \cellcolor[HTML]{C0C0C0}66.67         & \cellcolor[HTML]{C0C0C0}\textbf{84.62}   & \cellcolor[HTML]{C0C0C0}100 & \cellcolor[HTML]{C0C0C0}\textbf{75.00} & \cellcolor[HTML]{C0C0C0}\textbf{88.24} & \cellcolor[HTML]{C0C0C0}66.67        & \cellcolor[HTML]{C0C0C0}83.33          & 0      & \cellcolor[HTML]{C0C0C0}56.25           & \cellcolor[HTML]{C0C0C0}\textbf{20.00} & \cellcolor[HTML]{C0C0C0}\textbf{26.67} & \cellcolor[HTML]{C0C0C0}\textbf{88.24} & \cellcolor[HTML]{C0C0C0}\textbf{66.67} & \cellcolor[HTML]{C0C0C0}\textbf{61.54} & \cellcolor[HTML]{C0C0C0}66.67        \\
                            & 50.73                                  & \cellcolor[HTML]{C0C0C0}\textbf{0}     & \cellcolor[HTML]{C0C0C0}\textbf{0}   & \cellcolor[HTML]{C0C0C0}\textbf{0}     & 14.19   & \cellcolor[HTML]{C0C0C0}\textbf{0}     & \cellcolor[HTML]{C0C0C0}\textbf{0.0017} & \cellcolor[HTML]{C0C0C0}\textbf{0.00050} & \cellcolor[HTML]{C0C0C0}0              & \cellcolor[HTML]{C0C0C0}0             & \cellcolor[HTML]{C0C0C0}\textbf{0.00025} & \cellcolor[HTML]{C0C0C0}0   & \cellcolor[HTML]{C0C0C0}\textbf{0}     & \cellcolor[HTML]{C0C0C0}\textbf{0}     & \cellcolor[HTML]{C0C0C0}0            & \cellcolor[HTML]{C0C0C0}0.00075        & 0      & \cellcolor[HTML]{C0C0C0}0.0025          & \cellcolor[HTML]{C0C0C0}\textbf{0}     & \cellcolor[HTML]{C0C0C0}\textbf{0}     & \cellcolor[HTML]{C0C0C0}\textbf{0}     & \cellcolor[HTML]{C0C0C0}\textbf{0}     & \cellcolor[HTML]{C0C0C0}\textbf{0}     & \cellcolor[HTML]{C0C0C0}0            \\
\multirow{-3}{*}{\textbf{AutoYara}}  & 0.9                                    & \cellcolor[HTML]{C0C0C0}\textbf{1.3}   & \cellcolor[HTML]{C0C0C0}\textbf{1.5} & \cellcolor[HTML]{C0C0C0}\textbf{1.5}   & 0.9     & \cellcolor[HTML]{C0C0C0}\textbf{7.6}   & \cellcolor[HTML]{C0C0C0}\textbf{1.0}    & \cellcolor[HTML]{C0C0C0}\textbf{0.9}     & \cellcolor[HTML]{C0C0C0}1.0            & \cellcolor[HTML]{C0C0C0}1.5           & \cellcolor[HTML]{C0C0C0}\textbf{0.7}     & \cellcolor[HTML]{C0C0C0}1.0 & \cellcolor[HTML]{C0C0C0}\textbf{1.5}   & \cellcolor[HTML]{C0C0C0}\textbf{2.6}   & \cellcolor[HTML]{C0C0C0}0.9          & \cellcolor[HTML]{C0C0C0}0.6            & 0.8    & \cellcolor[HTML]{C0C0C0}1.2             & \cellcolor[HTML]{C0C0C0}\textbf{1.2}   & \cellcolor[HTML]{C0C0C0}\textbf{0.8}   & \cellcolor[HTML]{C0C0C0}\textbf{6.0}   & \cellcolor[HTML]{C0C0C0}\textbf{1.9}   & \cellcolor[HTML]{C0C0C0}\textbf{0.4}   & \cellcolor[HTML]{C0C0C0}1.5          \\ \bottomrule
\end{tabular}
}
\end{table*}

To simulate this scenario, we use the Retro Hunt capability of VirusTotal\cite{Virustotal} (VT), combined with samples of malware shared by Twitter user @0xffff0800\footnote{\url{https://twitter.com/0xffff0800/status/1155876158740869121}}. For each family, we use AutoYara and VxSig to construct a Yara rule, and submit that rule to Retro Hunt. Retro Hunt will then return hits against that rule for \textit{all} executables submitted to VirusTotal within the last 90 days. With over one million new unique files submitted per day\footnote{\url{https://www.virustotal.com/en/statistics/}}, this allows us to get a better understanding about false-positive rates. The terms of our VT subscription limit the  number of queries we are allowed to run, and so only AutoYara and VxSig are evaluated in this section.

We submitted AutoYara rules for 14 malware families 
to VT, with the results shown in \autoref{tbl:retro_hunt}. Lines that begin with "N/A" indicate a failure to produce a rule, which only happened with VxSig. Note that in each of these cases, AutoYara was able to produce a rule with 100\% TP rate. This may indicate a strength to our approach allowing for signature construction from all portions of the file. 

In many cases VT returns only a few hits, but some rules return hundreds if not thousands of hits. Due to the time-intensive nature of analyzing these results, we only review up to 50 returned hits to estimate the results. We can see that AutoYara achieves a 100\% TP rate on 11 out of the 20 families, requiring no human interaction at all. This is noteworthy due to the fact that these rules are run over 90 million files, indicating our biclustering approach obtains exceptionally low false-positive rates. 
In 15 of the 20 cases AutoYara performed better or had identical results as VxSig. In half of the instances where VxSig performed better, the results were a marginal improvement but still did not obtain any TP hits (APT26 and CobaltGroup). 

There are some failure cases, of varying degrees of severity. The Shamoon malware did not fire on any new samples, so no results were returned. In this case we learn nothing from the rule, but no analyst time was wasted on false positives. The Sofacy malware returns 60\% malware, though most of which appears to be from other families. This result is still useful in the hunt situation, but the FP rate on benign applications is still high. The Petya rule generated had a significant false positive issue, and would not be informative in practice.

Overall these results are encouraging, and indicate that AutoYara can be useful for hunt activities when a limited number of examples of malware may be available. We are able to generate useful rules in the majority of cases, using a small number of samples, often with no false positives over 90 million files.

\subsection{Industry Human Comparison Testing} \label{sec:autoyara_vs_human}

Our last test is based on real-world work performed by professional malware researchers. As part of day-to-day operations, a need to develop Yara rules to detect known and particularly challenging families exists. Three malware analysts were asked to develop Yara rules for 24 malware families. The stated goal was to provide maximal coverage for malware samples with minimal false-positive rate. While there is some tolerance for cross-tagging other malicious families with a Yara rule intended for a single family, a false positive rate of $\geq 0.1\%$ on benign files was unacceptable.

The results with TP and FP rates are shown in \autoref{tbl:bot_v_human}, where Analysts A and B are both considered experts with multiple years of experience. Analyst B used YarGen as part of their standard workflow, and found it was insufficient on its own in all cases --- so Analyst B's results subsume YarGen. Several rows of the table are empty because neither analyst was able to complete all their work, and new priorities eventually subsumed these and were never completed. This high demand on their time, and its time intensive nature, is part of the problem we are trying to solve. For example, Analyst B spent 78 minutes on the baldr family, but did not create a usable rule due to a false-positive rate of 11.8\%. Analyst C was not a part of the business workflow, and so was able to spend two weeks generating signatures for all 24 families. 

On this production data, the results of AutoYara are significant. For 21/24 families, AutoYara successfully produced useful rules with reasonable TP rates and exceptionally low FP rates, either obtaining exactly 0 FPs on over 400,000 test files, or extremely low rates such as 0.00025\%. While AutoYara usually performed slightly worse than analysts, it produced better results than one or more analysts on 10/24 of the families (bkff, ertfor, firefly, navattle, nezchi, olympicdestroyer, phds, sekur, subroate, and wuca). Based on these results, Analyst B could have saved 44\% of the time they spent working on families that AutoYara was able to capture, and instead focused on the more difficult samples like baldr, darkvnc, or plurox. Similarly, Analyst C could have saved 86\% and Analyst A could have saved 100\% of their time, and could have instead focused on these more challenging cases that did not avail to automation. 

In comparison, the Greedy approach used by many prior works was only able to produce usable rules on 5/24 families, and only on the easiest samples. VxSig was able to produce rules for 10/24 families, only half of AutoYara. Note that in only two instances did VxSig outperform an analyst, and it requires up to \textit{7.9 hours} to process a single family. This significantly hampers VxSig's usability. 

\begin{wraptable}{r}{5.25cm}
\caption{
AutoYara results improved by an inexperienced user with one attempt at editing the produced rule. 
}
\label{tbl:autoyara_human_join}
\begin{tabular}{@{}lccc@{}}
\toprule
           & \multicolumn{1}{l}{Time}     & \multicolumn{2}{l}{Human+AutoYara} \\ \cmidrule(l){3-4} 
Family     & \multicolumn{1}{l}{(min)} & TP\%             & FP\%            \\ \midrule
baofa      & 2                        & 50               & 0               \\
darkvnc    & 1                        & 50               & 0.83            \\
jongiti    & 2                        & 70.59            & 0               \\
ladyoffice & 1                        & 100              & 0               \\
subroate   & 2                        & 46.67            & 0               \\
zcash      & 1                        & 100              & 0               \\ \bottomrule
\end{tabular}
\end{wraptable}

Another benefit to AutoYara is that it is easy for an Analyst to modify the rule to improve the performance. In \autoref{tbl:autoyara_human_join} we show the results where a user with only one course on reverse engineering, and no professional experience developing rules, attempted to modify the rules produced by AutoYara. For these 6 families they were able to improve the TP/FP rate, requiring only a few minutes for each file.

\section{When Will AutoYara Work Well?} \label{sec:autoyara_depth}

We now investigate some of the reasons why/when AutoYara will work well. First we will show that structural consistency in byte strings across samples is related to AutoYara's performance, which is expected. Diving deeper, we investigate the generated rules to better understand the type of content AutoYara uses, how the generated rules perform, and how the performance compares to a domain expert's results. 

\subsection{Byte Similarity Investigation}
To help better understand when and why AutoYara performs well, we performed an investigation of the similarity between malware samples using SSDEEP\cite{Kornblum2006}. SSDEEP creates a similarity digest from the raw bytes of the input files, and can return a score in the range of 0 to 100. In general, any score $\geq $ 20 is a "match", and scores quickly drop off to 0 for non-similar content. While SSDEEP comparisons are heuristic, we found visualizing the connected components of malware families based on their pairwise SSDEEP scores to be a useful diagnostic.

\begin{figure}[!ht]
\centering

\subfloat[Olympic Destroyer]{
    \adjustbox{max width=0.5\columnwidth}{%
    \includegraphics[]{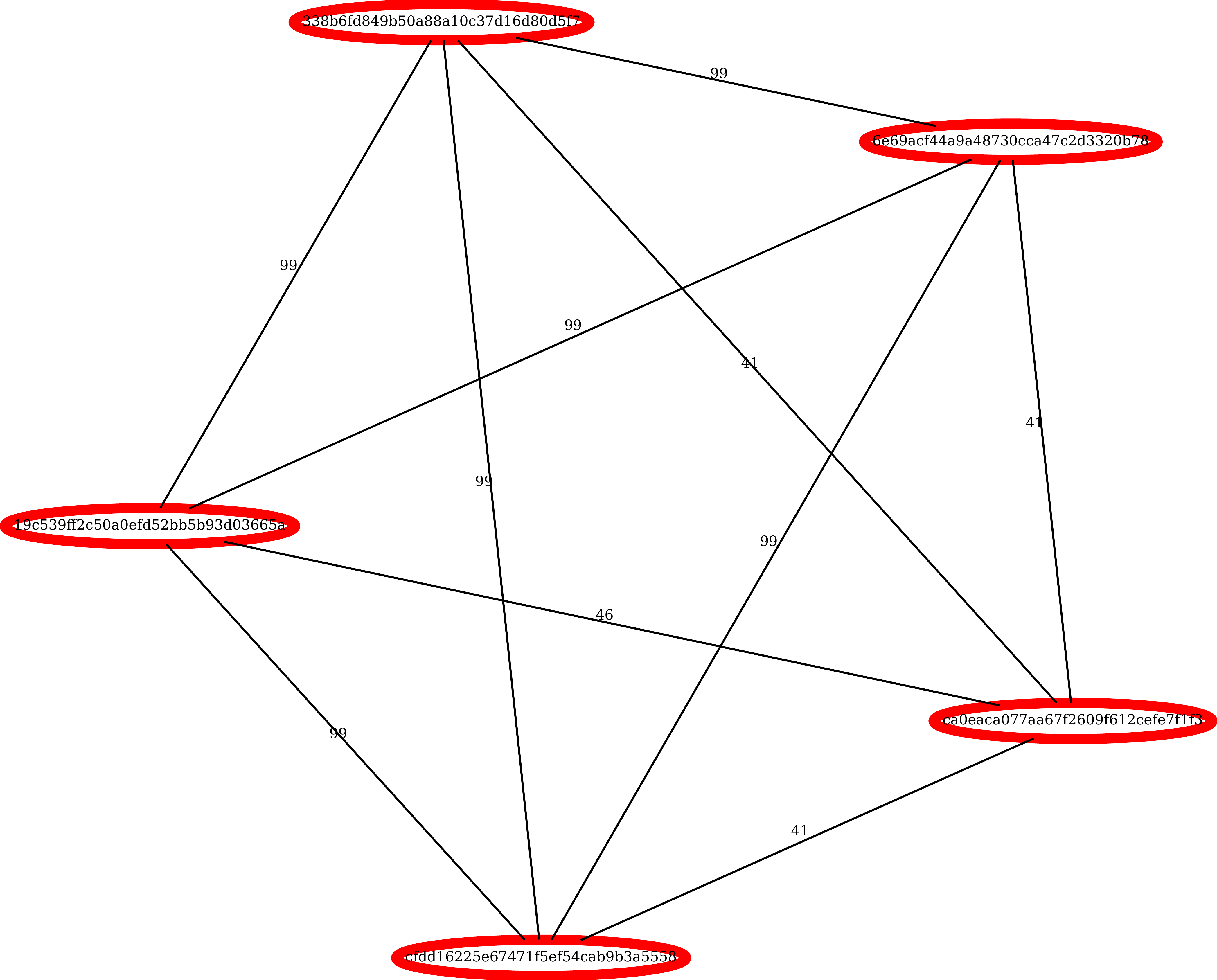}
\label{fig:olympic_graph}
    }
}
\subfloat[Dragonmess]{
    \adjustbox{max width=0.5\columnwidth}{%
    \includegraphics[]{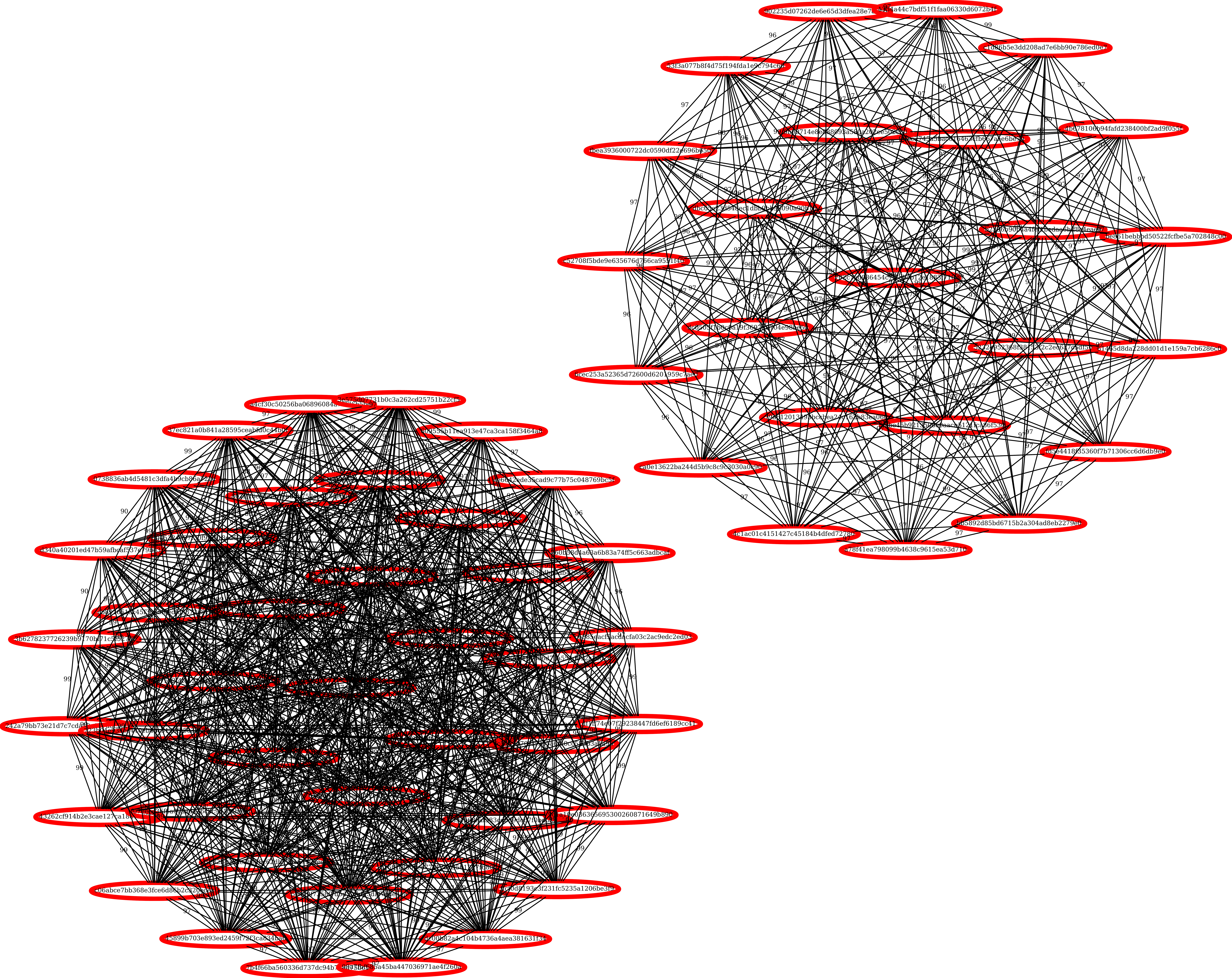}
\label{fig:dragonmess_graph}
    }
}
\caption{Connected components graph based on SSDEEP similarities for two families where AutoYara did well.}
\label{fig:test}
\end{figure}

For example, in Figure \autoref{fig:olympic_graph} and Figure \autoref{fig:dragonmess_graph} we see the graphs created for the Olympic Destroyer and Dragonmess malware families. Both families exhibit clustering into a few densely connected sub-graphs. Unsurprisingly, AutoYara rules perform very well on these families. The intuition is that the high byte similarity of these inputs make for convenient rules.  These results support the claim that a biclustering based approach to represent a single family is fruitful. The data itself tends to lead to natural sub-family populations which are easier to represent with the biclustering process. 

We found that AutoYara had a higher failure rate for the AVClass-labeled corpus, compared to the production data. We suspect that this was caused by a larger amount of label noise in those tests, since the AVClass tool is not perfect, and depends on the output from several AV products --- which are also not perfect. This accumulation of error could have caused increased noise, making any attempts at rule construction difficult.

\begin{figure}[!ht]
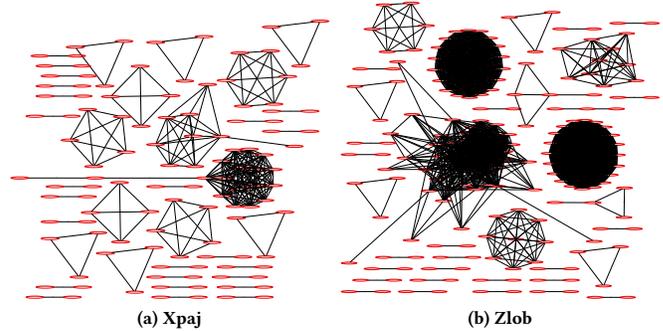

\centering
\subfloat[Xpaj]{
    \adjustbox{max width=0.5\columnwidth}{%
    \includegraphics[]{figs_new/xpaj_1000.pdf}
\label{fig:xpaj}
    }
}
\subfloat[Zlob]{
    \adjustbox{max width=0.5\columnwidth}{%
    \includegraphics[]{figs_new/zlob_1000.pdf}
\label{fig:zlob}
    }
}
\caption{Families where AutoYara did not perform well.}
\label{fig:test2}
\end{figure}

The graphs in Figure \autoref{fig:xpaj} and Figure \autoref{fig:zlob} support this hypothesis, showing much greater disparity in pairwise similarity of files from the Xpaj and Zlob families that were most challenging for rules produced by AutoYara. While there are dominant clusters, the presence of a large number of small tuples indicates a potentially noisy input set that would thus reduce the effectiveness of any tool attempting to produce a useful rule. 

\begin{figure}[!ht]
\centering
\subfloat[AutoYara-generated rules]{
    \adjustbox{max width=\columnwidth}{%
    \includegraphics[]{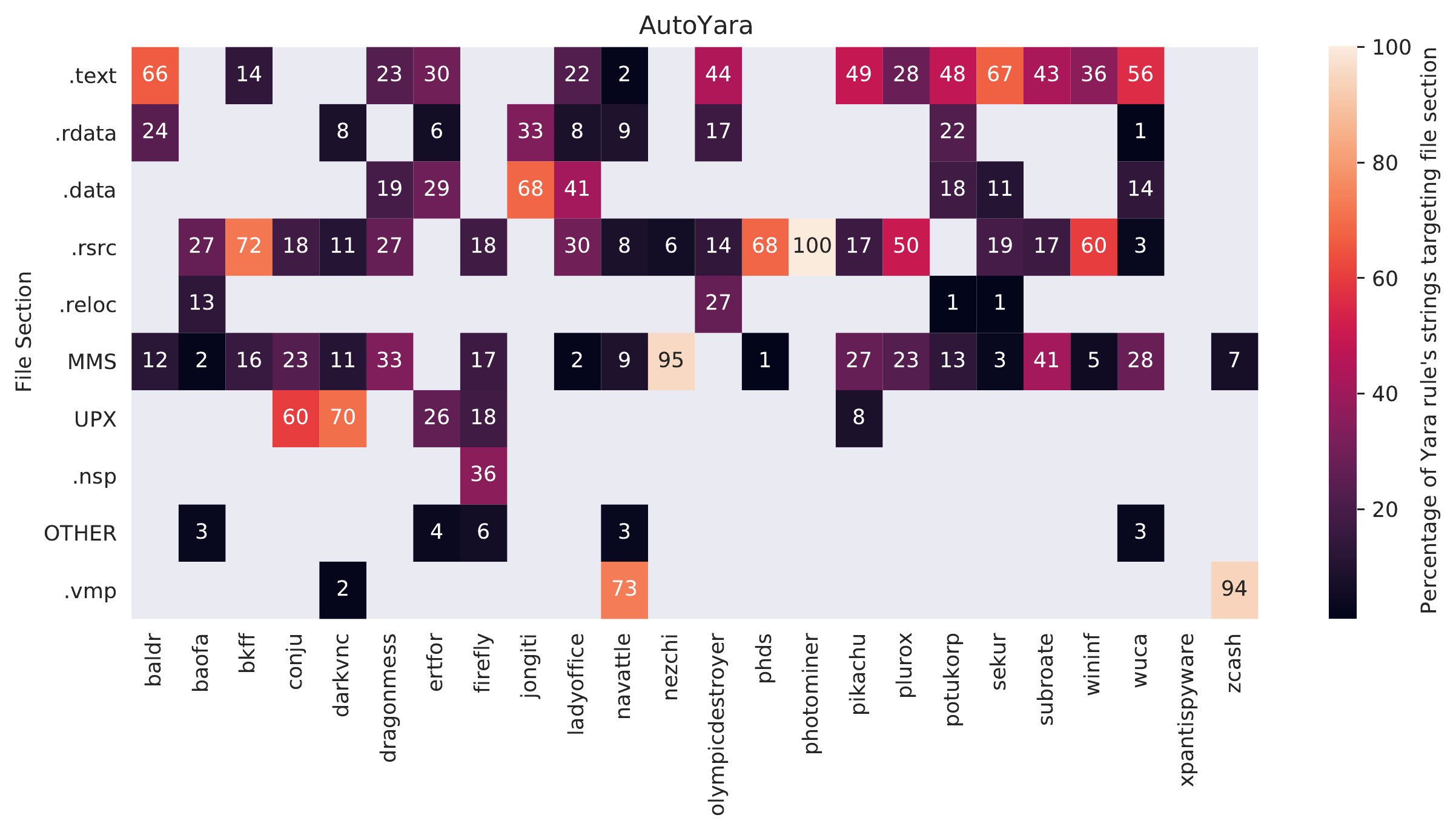}
\label{fig:heatgraph_autoyara}
    }
}
\newline
\subfloat[Manually-generated rules]{
    \adjustbox{max width=\columnwidth}{%
    \includegraphics[]{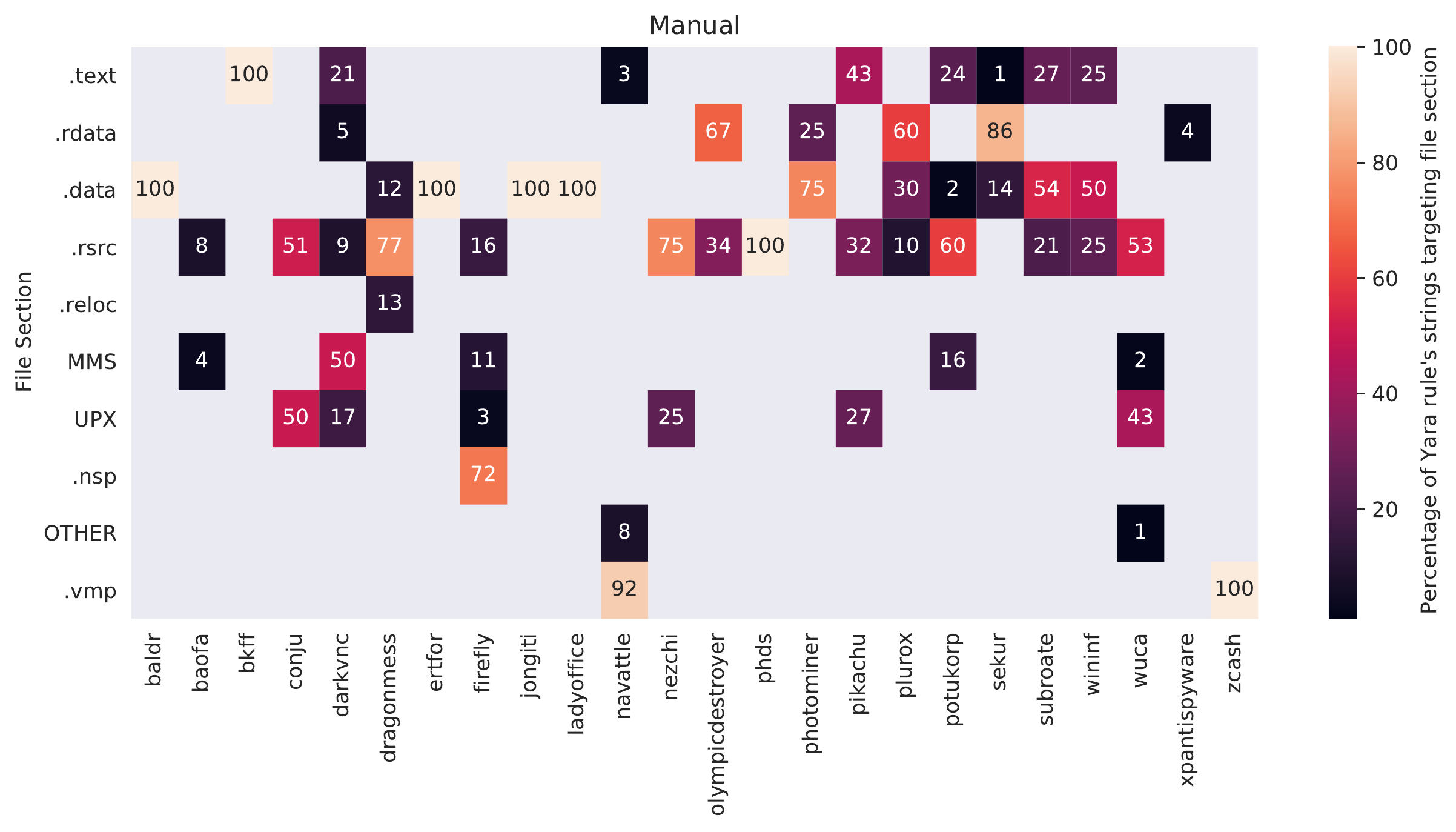}
\label{fig:heatgraph_manual}
    }
}
\caption{File sections targeted by AutoYara and analyst show that both tend to use similar section to generate rules. AutoYara tends to target the \texttt{.text} section, while analyst tend to target the .data section. "MMS" describes rules that were in multiple file sections; "OTHER" corresponds condensed file sections that were not relevant to our analysis.}
\label{fig:test3}
\end{figure}

\subsection{Reverse Engineering Investigation}
In addition, we conducted reverse engineering of the rules that AutoYara created. We looked at rules generated for the Baldr and Conju family in detail, and noticed that all the main file section were targeted. Rules targeting the resource sections were strings belonging to the application manifest and DLL names, rules in the text sections pointed to regular functions blocks, rules in the data section were long strings not easily interpretable, and in the case of the Conju family many of the rules targeted the decompression stub of UPX. In general the content of the AutoYara rules were reasonable items to target. For code sections in particular, AutoYara did not tend to target the exact same functions, but the functions selected seemed reasonably specific of the sample. 

We also analyzed the different file sections each of the rules generated by AutoYara and Analyst C  (the only analyst to investigate all malware families, hence we use their results). Figure \autoref{fig:heatgraph_autoyara} and Figure \autoref{fig:heatgraph_manual} shows that both AutoYara and our analyst wrote rules targeting the \texttt{.text}, \texttt{.rdata}, \texttt{.data}, \texttt{.rsrc}, and \texttt{.reloc} sections. We also noticed that when AutoYara generated rules, it had a tendency to target the \texttt{.text} section, whereas our analyst tended to write rules targeting the data section. It's likely that AutoYara targets the \texttt{.text} section due to its high entropy and because it can find blocks of instructions that are shared across multiple binaries. On the other hand, analysts tend to use the data section more because it contains globally accessible or predefined data such as strings or constants that are easier to identify, extract, and understand. The tendency of AutoYara to target high-entropy areas can also contribute to its small bias towards the \texttt{.UPX1} sections of UPX-packed binaries (shown as positive values on the UPX row in Figure \autoref{fig:heatgraph_autoyara}). However, further exploration of this subject is needed to determine if there are strong correlations, or if this is an artifact of our small sample size. Due to the high cost of such experiments such a larger manual study may be difficult to perform. We are aware of no other work that has compared manual-vs-automatic generated rules.

Finally, we look at the commonality in rule behavior, with results in \autoref{tbl:mean_memory_segments}. $\bar{X}$ indicates the mean number of file sections (e.g., $\bar{X}=2$ if \texttt{.text} and \texttt{.data} are represented in the rule) that a rule component (each string within a Yara rule) hit upon in the test data. Using this we can see that AutoYara and manually built rules have a high correlation, both tending to have strings that hit only one file section, or multiple section, indicating that a degree of similarity in the types of content used. 

The Similarity column indicates the overlap in the test set for what executables the final rule triggered on. A similarity of 100\% means that AutoYara and the human analyst's intersection is perfect, and 0\% indicates no overlap in the files flagged. From these results we see that again, AutoYara and manual domain expert constructed rules tend to agree upon and find similar files in the test set, with most differences caused by differing false-positive rates (e.g., ladyoffice, where AutoYara has only 2 hits that are TPs, and the analyst hits only the 3 TPs).

\begin{table}
    \caption{Summary of the $\overline{X}$ (mean) number of file section each of the strings in a rule targeted, and the percentage similarity between the matches generated by each set of rules. Similarity was calculated as the number of common hits over all the unique hits both rules had.}
    \label{tbl:mean_memory_segments}
    \begin{tabular}{lccccr}
    \toprule
    {} & \multicolumn{2}{c}{AutoYara} & \multicolumn{2}{c}{Manual} &  \\
    \cmidrule(r){2-3} \cmidrule(r){4-5}
    Family & $\overline{X}$ & Hits & $\overline{X}$ & Hits & \% Similarity \\
    \midrule
    baldr            &     1.00 &    6 &   1.00 &   14 &       43 \\
    baofa            &     1.99 &   51 &   2.00 &   53 &       37 \\
    bkff             &     2.63 &  103 &   1.00 &   41 &       40 \\
    conju            &     1.00 &  149 &   1.00 &  153 &       97 \\
    darkvnc          &     3.00 &   71 &   2.20 &    5 &       01 \\
    dragonmess       &     1.03 &   77 &   3.00 &  100 &       57 \\
    ertfor           &     2.57 &   35 &   1.00 &    1 &       00 \\
    firefly          &     1.67 &   17 &   3.50 &   53 &       41 \\
    jongiti          &     1.00 &  107 &   1.00 &  154 &       69 \\
    ladyoffice       &     1.00 &   24 &   1.00 &   31 &       77 \\
    navattle         &     2.00 &   49 &   4.00 &   45 &       84 \\
    nezchi           &     1.00 &   65 &   1.00 &   70 &       93 \\
    olympicdestroyer &     1.00 &   57 &   1.00 &   58 &       98 \\
    phds             &     1.00 &  144 &   1.00 &  148 &       90 \\
    photominer       &     1.00 &   24 &   1.00 &   28 &       86 \\
    pikachu          &     1.17 &   38 &   1.67 &   50 &       69 \\
    plurox           &     1.00 &   11 &   1.00 &   12 &       92 \\
    potukorp         &     2.75 &  106 &   2.75 &  197 &       75 \\
    sekur            &     1.33 &   25 &   1.71 &   29 &       80 \\
    subroate         &     1.00 &   94 &   1.00 &  207 &       28 \\
    wininf           &     1.00 &  157 &   1.00 &  157 &       100 \\
    wuca             &     4.11 &   63 &   3.50 &  310 &       12 \\
    xpantispyware    &     1.00 &   70 &   1.50 &   79 &       89 \\
    zcash            &     1.65 &   30 &   1.00 &   34 &       83 \\
    \bottomrule
    \end{tabular}
\end{table}

\section{Conclusion} \label{sec:conclusion}

Using large byte-based n-grams combined with a new biclustering algorithm, we have developed AutoYara, a new approach to automatically constructing Yara rules from an example set of malware families. In many cases AutoYara can perform as well as an expert analyst, and production testing indicates it could save an analyst over 44\% of their time spent making Yara rules. This can allow analysts to get through more of their workload, and spend human effort on the most challenging cases that are currently beyond automation.

\bibliographystyle{ACM-Reference-Format}
\bibliography{ref,Mendeley}

\begin{appendices}

\section{Extra Nuance Details of AutoYara}

As presented, the version of AutoYara described will be effective and matches the approach our implementation takes. We make note of small enhancements that we found improved the reliability of our results. First, rather than picking just scale or bistochastic based normalization, we use both approaches. We run each individually, create a rule, and select the best rule according to the input set coverage as described in \autoref{algo:autoyara}. Statistical testing has found there is no significant difference in the performance of each approach, so one is not uniformly better than the other. But on individual cases, one may perform better. 

A second detail that was tested, but not included, is to run the Variational GMM multiple times. We found this can be helpful as the VGMM approach is not deterministic, and sometimes converges to local optima that do not perform well. However this is expensive. Instead, if clustering fails we fall back to using HDBSCAN on the inputs, selecting "biclusters" as the returned clusters constrained to the features that occurred in more than 50\% of the clusters. This again was not necessary for our approach to work, but a micro-optimization we found improved the reliability of our results.



\section{VxSig Approach } \label{sec:vxsig}

For completeness on the VxSig approach and reproducibility, we will describe how we used it and the results here. VxSig requires a tool called BinDiff, which is used to compare the similarity of assembly sequences and their control flow graph. We used VxSig with Ghidra\footnote{\url{https://ghidra-sre.org/}}, a reverse engineering tool developed by the NSA. Ghidra performs the disassembly of the executables, exporting the results to a "BinExport" format, which is used by BinDiff to find common assembly blocks between multiple programs. VxSig uses this result to create a rule from the largest common sub-sequences of assembly instructions shared by all of the inputs. This is the general approach described by \textcite{Blichmann2008}. 

In using VxSig it is possible to set it up to run in a completely automated fashion, disassemble each sample from a family, run BinDiff, and produce the Yara Rule. In practice, we found that this almost always resulted in a failure to build a rule (i.e, the tool VxSig itself would throw an error as it was unable to find any common sub-sequence) or produce a degenerate rule that matched all or no executables. 
As such, VxSig was only able to run successfully on the jongiti in a fully automated fashion. To get results for all other families, we followed the below process:

\begin{enumerate}
    \item Attempt to disassemble all files
    \item Compute the average number of disassembled bytes, and discard any sample that had more than 1 standard deviation more or less bytes (i.e., too many or too few instructions)
    \item Run BinDiff and attempt to produce a rule. 
    \item If VxSig still fails, try VxSig on only each pair of inputs --- and find the pairs that produce good rules. 
    \item Run VxSig on only the found pairs that worked individually. If it still fails, pick the pair of inputs that had the largest similarity. 
\end{enumerate}

The primary issue is VxSig's requirement to match \textit{all} inputs, which is not reliable. Many malware samples fail to disassemble without analyst intervention, which can make VxSig fail due to having no input to operate on. This occurred with the bkff, APT36, GrandCrab, and WannaCry samples. A larger issue is that malware samples can be intentionally malformed, and disassembly may run "successfully" but produce a bad disassembly sequence. In this case, it becomes impossible for VxSig to match the badly parsed file to other inputs. This is common as disassembly is a non-trivial task, in which malware authors intentionally attempt to thwart or at least slow the work of analysts. 

That said, when VxSig works, it can be quite powerful. It was the only method to produce a usable rule for the baldr family, and produced better rules than AutoYara in 6 cases for our production data. However, in most cases VxSig fails to match new samples of the family and has higher false-positive rates than AutoYara. We see VxSig as being complementary, and a tool to be used by analysts on the most difficult malware families that tools like AutoYara can not handle. These instances already require manual intervention, and the work an analyst may do as part of their normal process to reverse engineer the samples (to find something to make a signature out of) will aid in building a more effective signature from VxSig.

\end{appendices}

\end{document}